\documentclass[fleqn,usenatbib]{mnras}

\usepackage{newtxtext,newtxmath}

\usepackage[T1]{fontenc}

\DeclareRobustCommand{\VAN}[3]{#2}
\let\VANthebibliography\thebibliography
\def\thebibliography{\DeclareRobustCommand{\VAN}[3]{##3}\VANthebibliography}

\usepackage{graphicx}	
\usepackage{amsmath}	
\usepackage{anyfontsize} 
\usepackage{longtable}
\usepackage{orcidlink}
\usepackage[dvipsnames]{xcolor}
\usepackage{dashrule}

\title[The Type~Ibn/Icn SN~2024abvb]{SN~2024abvb: a Type~Ibn/Icn supernova with evidence of helium and an extreme offset from its host galaxy}

\author[C. Aster et al.]{Callum Aster$^{1}$\thanks{E-mail: aubreyc@cardiff.ac.uk}\,\orcidlink{0009-0006-6872-3725},
Cosimo Inserra$^{1}$\orcidlink{0000-0002-3968-4409},
Andrea Pastorello$^{2}$\orcidlink{0000-0002-7259-4624},
Joseph P Anderson$^{3}$\orcidlink{0000-0003-0227-3451},
Franz Erik Bauer$^{4}$\orcidlink{0000-0002-8686-8737},
\newauthor
K.\ Azalee Bostroem$^{5,6}$\orcidlink{0000-0002-4924-444X}
Kenneth C. Chambers$^{7}$\orcidlink{0000-0001-6965-7789},
Ting-Wan Chen$^{8}$\orcidlink{0000-0002-1066-6098},
Joseph R. Farah$^{9,10}$\orcidlink{0000-0003-4914-5625},
\newauthor
Morgan Fraser$^{11}$\orcidlink{0000-0003-2191-1674},
Dino Pierluigi Fugazza$^{12}$\orcidlink{0000-0001-9942-277X},
Mariusz Gromadzki$^{13}$\orcidlink{0000-0002-1650-1518},
Claudia P. Guti\'errez$^{14,15}$\orcidlink{0000-0003-2375-2064},
\newauthor
D. Andrew Howell$^{9,10}$,
Erkki Kankare$^{16}$\orcidlink{0000-0001-8257-3512},
Tom L. Killestein$^{17}$\orcidlink{0000-0002-0440-9597},
Niilo Koivisto$^{16}$\orcidlink{0009-0007-7151-7313},
Giorgos Leloudas$^{18}$\orcidlink{0000-0002-8597-0756},
\newauthor
J. D. Lyman$^{17}$\orcidlink{0000-0002-3464-0642}
Kyle Medler$^{7}$\orcidlink{0000-0001-7186-105X},
Shane Moran$^{19}$\orcidlink{0000-0001-5221-0243},
Tomás E. Müller-Bravo$^{20,21}$\orcidlink{0000-0003-3939-7167},
Giuliano Pignata$^{4}$\orcidlink{0000-0003-0006-0188},
\newauthor
Miika Pursiainen$^{17}$\orcidlink{0000-0003-4663-4300},
Fabio Ragosta$^{22,23}$\orcidlink{0000-0003-2132-3610},
Andrea Reguitti$^{2,12}$\orcidlink{0000-0003-4254-2724},
Jesper Sollerman$^{24}$\orcidlink{0000-0003-1546-6615},
Giorgio Valerin$^{2}$\orcidlink{0000-0002-3334-4585},
\newauthor
Ben Warwick$^{17}$\orcidlink{0009-0005-8379-3871},
David R. Young$^{25}$\orcidlink{0000–0002–1229–2499}.
\\
$^{1}$Cardiff Hub for Astrophysics Research and Technology, School of Physics \& Astronomy, Cardiff University, Queens Buildings, The Parade, Cardiff, CF24 3AA, UK\\
$^{2}$INAF-Osservatorio Astronomico di Padova, Vicolo dell’Osservatorio 5, 35122 Padova, Italy\\
$^{3}$European Southern Observatory, Alonso de Córdova 3107, Vitacura, Casilla 19001, Santiago, Chile\\
$^{4}$Instituto de Alta Investigaci{\'{o}}n, Universidad de Tarapac{\'{a}}, Casilla 7D, Arica, 1010000, Chile\\
$^{5}$Steward Observatory, University of Arizona, 933 North Cherry Avenue, Tucson, AZ 85721-0065, USA\\
$^{6}$LSST-DA Catalyst Fellow\\
$^{7}$Institute for Astronomy, University of Hawaii at Manoa, 2680 Woodlawn Dr., Hawaii, HI 96822, USA\\
$^{8}$Graduate Institute of Astronomy, National Central University, 300 Jhongda Road, 32001 Jhongli, Taiwan\\
$^{9}$Las Cumbres Observatory, 6740 Cortona Drive, Suite 102, Goleta, 93117-5575, CA, USA\\
$^{10}$Department of Physics, University of California, Santa Barbara, 93106-9350, CA, USA\\
$^{11}$School of Physics, University College Dublin, LMI Main Building, Beech Hill Road, Dublin 4, D04 P7W1\\
$^{12}$INAF – Osservatorio Astronomico di Brera, Via E. Bianchi 46, I23807 Merate (LC), Italy\\
$^{13}$Astronomical Observatory, University of Warsaw, Al. Ujazdowskie 4, 00-478 Warszawa, Poland\\
$^{14}$Institut d'Estudis Espacials de Catalunya (IEEC), Edifici RDIT, Campus UPC, 08860 Castelldefels (Barcelona), Spain\\
$^{15}$Institute of Space Sciences (ICE, CSIC), Campus UAB, Carrer de Can Magrans, s/n, E-08193 Barcelona, Spain\\
$^{16}$Department of Physics and Astronomy, University of Turku, 20014 Turku, Finland\\
$^{17}$Department of Physics, University of Warwick, Gibbet Hill Road, Coventry CV4 7AL, UK\\
$^{18}$DTU Space, Department of Space Research and Space Technology, Technical University of Denmark, Elektrovej 327, 2800 Kgs. Lyngby, Denmark\\
$^{19}$School of Physics and Astronomy, University of Leicester, University Road, Leicester LE1 7RH, UK\\
$^{20}$School of Physics, Trinity College Dublin, The University of Dublin, Dublin 2, Ireland\\
$^{21}$Instituto de Ciencias Exactas y Naturales (ICEN), Universidad Arturo Prat, Chile\\
$^{22}$Dipartimento di Fisica ‘Ettore Pancini’, Universita` di Napoli Federico II, Via Cinthia 9, 80126 Naples, Italy\\
$^{23}$INAF – Osservatorio Astronomico di Capodimonte, Via Moiariello 16, I-80131 Naples, Italy\\
$^{24}$The Oskar Klein Centre, Department of Astronomy, Stockholm University, AlbaNova SE-10691, Stockholm, Sweden\\
$^{25}$Astrophysics Research Centre, School of Mathematics and Physics, Queen’s University Belfast, Belfast BT7 1NN, UK\\
}
\date{Accepted XXX. Received YYY; in original form ZZZ}

\pubyear{\the\year{}}

\begin{document}
\label{firstpage}
\pagerange{\pageref{firstpage}--\pageref{lastpage}}
\maketitle

\begin{abstract}
\label{sec:abstract}

We present spectroscopic and photometric observations and analysis of SN~2024abvb, a peculiar transitional Type~Ibn/Icn supernova located at an unusually large projected distance from its host galaxy (21.5~kpc). SN~2024abvb displays an extended rise time in the $g$- and $o$-bands (10.1 and 10.6~days respectively), followed by a linear decline in all photometric bands.  Comparisons with other supernova subclasses show that the photometric and spectroscopic evolution of SN~2024abvb are distinct from Type~Ibn and Type~Icn events, with a higher peak $r$-band luminosity and lower blackbody temperatures. Spectra reveal an initial blue continuum and narrow P-Cygni profiles, with C~{\sc ii} $\lambda$5890 dominating in emission, persisting at late phases, and showing a rapid decline in the expansion velocity. Weak He~{\sc i} $\lambda$5876 features are tentatively detected at early times. Analysis of progenitor scenarios rules out thermonuclear origins based on incompatible light curve shapes and spectral signatures. A rare massive star progenitor appears unlikely given the low local star formation rate. The most plausible origin is an ultra-stripped supernova scenario involving a binary system; this best explains the observed separation from the host, the low circumstellar material mass, the fast photometric evolution and the low nickel production, although a discrepancy in model versus observed ejecta mass remains. These results reinforce the classification of SN~2024abvb as a distinctive Type~Ibn/Icn event and highlight the diversity of progenitor channels for interacting supernovae.
\end{abstract}

\begin{keywords}
transients: supernovae -- supernovae: individual: SN~2024abvb
\end{keywords}



\section{Introduction}
\label{sec:intro}

Massive stars, those with an initial mass at the Zero Age Main Sequence exceeding roughly 8~$M_{\odot}$, end their lives as core-collapse supernovae (CCSNe).  Observationally, CCSNe divide into hydrogen‐rich (Type~II) and hydrogen‐poor (Type~I) classes based on the presence or absence of Balmer lines in their spectra \citep{1941PASP...53..224M,1997ARA&A..35..309F}. This spectroscopic dichotomy reflects fundamentally different pre‐explosion evolutionary pathways, driven by mass‐loss processes.

Hydrogen‐poor SNe (Type~I) can arise when strong stellar winds or eruptive episodes remove the progenitor’s hydrogen envelope.  In isolated massive stars, radiatively driven winds become increasingly efficient at higher metallicity and luminosity, eventually peeling away the outer layers to reveal helium‐ and heavier‐element cores \citep{2000ARA&A..38..613K}.  Such progenitors give rise to ``stripped‐envelope'' supernovae (SE~SNe), subdivided into Type~Ib (helium lines present), Type~IIb (hydrogen mostly stripped) and Type~Ic (helium lines weak or absent) events \citep{1995ApJ...448..315W}.  Binary interactions, through Roche‐lobe overflow or common‐envelope ejection, can produce analogous stripping even for stars whose winds alone would not suffice \citep[e.g.][]{Smith_2011}.

Wolf–Rayet (WR) stars, characterised by their broad emission‐line spectra, constitute one of the possible SE~SN progenitors \citep{Sukhbold_2016}. Within the WR family, nitrogen‐rich (WN) stars are generally linked to Type~Ib explosions, whereas carbon‐rich (WC and WO) stars correspond to Type~Ic events \citep{1985ApJ...294L..17W, 2009A&A...508..371H}.  However, lower‐mass helium stars, produced via binary mass transfer, may also explode as SE~SNe \citep[e.g.][]{2014ApJ...794...23D, 2016MNRAS.457..328L, 2018MNRAS.478.4162P}, complicating the mapping between spectral subtype and progenitor mass \citep{2013MNRAS.436..774E}.  

When significant amounts of circumstellar material (CSM) remain close to the progenitor at collapse, either retained from previous eruptions or not fully dispersed by winds, the SN ejecta can collide with this CSM. The interaction converts kinetic energy into radiation, and the observed SN show multi-component emission profiles generated by the ejecta and the CSM.
These interacting SNe are spectroscopically classified as Type~IIn (hydrogen-rich CSM), Type~Ibn (helium-rich CSM; \citealp{2007Natur.447..829P,2007ApJ...657L.105F}), Type~Icn (carbon‐rich CSM; \citealp{2022Natur.601..201G}) or Type~Ien (silicon- and sulfur-rich CSM; \citealp{2025Natur.644..634S}).  Type~Ibn SNe exhibit strong, narrow He~{\sc i} lines and generally lack the oxygen and carbon signatures that characterize Type~Icn events, whereas Type~Icn SNe show prominent O~{\sc ii-iii} and C~{\sc ii-iii} lines alongside P-Cygni profiles at early times \citep{Pellegrino_2022}. The recently identified class of Type~Ien shows a lack of CNO but strong Si and S narrow emission lines \citep{2025Natur.644..634S}.
SN~2019hgp was the first firm Type~Icn classification \citep{2022Natur.601..201G} and has been followed by an additional four confirmed events: one reclassification (SN~2019jc; \citealp{Pellegrino_2022}) and three new discoveries (SNe~2021ckj, 2021csp, 2022ann; \citealp{2023A&A...673A..27N,Davis_2023, Perley_2022}).  These events share uniform characteristics: spectral line velocities of $\sim$\,10$^{3}$--10$^{4}\,\mathrm{km\,s^{-1}}$, consistent with WR progenitor winds; rise times to peak light of $\sim$\,7\, days; peak absolute magnitudes clustered around $M_{r}\approx -19$ mag; spectra dominated by narrow C and O emission/absorption, indicative of recently ejected carbon‐rich CSM.
The features shared among Type~Icn events, fast rise time and the presence of carbon, might indicate similar progenitor scenarios.  However, other photometric and spectroscopic properties suggest a wider diversity in progenitor scenarios.  These scenarios range from carbon‐rich, hydrogen‐ and helium‐depleted stars that underwent a major mass‐loss episode shortly before core collapse to compact objects in binary systems.

Here, we present the evolution of SN~2024abvb (\citealp[see also][]{hu2026sn2024abvbtypeicn,theintelcollaboration2026nestedasymmetrichhecircumstellar,2026arXiv260216227S}), a new member of the SN~Ibn/Icn family. 
SN~2024abvb was discovered by the Asteroid Terrestrial‐impact Last Alert System (ATLAS; \citealp{2018PASP..130f4505T}) on 22 November 2024 (MJD 60636.36, \citealp{2024TNSTR4579....1T}), and subsequently classified as a Type~Icn at $z=0.039$ by the NUTS collaboration (MJD 60641.90, \citealp{2024TNSCR4674....1S}). Due to the tentative detection of helium, we reconsider this classification. Type~Icn~SN are classified based on the absence of helium in their spectra. Thus, the tentative detection of helium in the medium-resolution spectra differs from a pure Type~Icn event classification, suggesting that a transitional classification is more appropriate. Here we present ultraviolet (UV) through near‐infrared (NIR) photometry alongside optical spectroscopy, with systematic comparisons to previously reported Type~Icn, Type~Ibn and transitional objects of Type~Ibn/Icn. Throughout the paper we assume a standard $\Lambda$CDM cosmology \citep{2011ApJS..192...18K} with $\Omega_{M}=0.27$, $\Omega_{\Lambda}=0.73$
and $H_{0}= 70\mathrm{km\,s^{-1}} \mathrm{Mpc^{-1}}$ which gives $D_{L}=$172.2~Mpc.

\begin{figure}
	\includegraphics[width=\columnwidth]{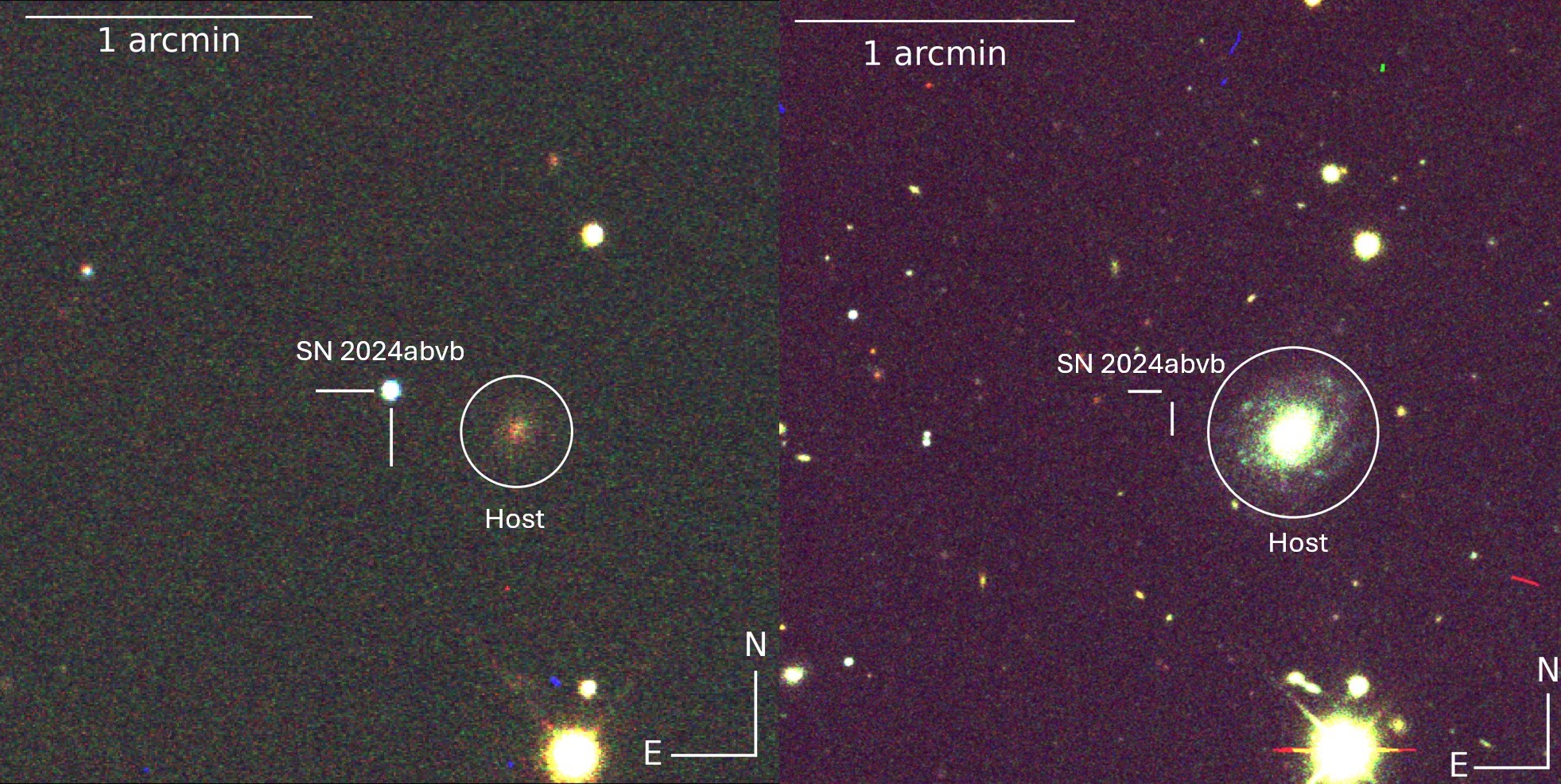}
    \caption{ \textbf{Left Panel:}$gri$ image stack from Las Cumbres Observatory (LCO) data (taken on 2024/11/28) with SN and host labelled. A 1 arcmin scale bar has been plotted at the top left of the image, with north oriented up and east to the left. \textbf{Right Panel:}$gri$ image stack from Dark Energy Spectroscopic Instrument (DESI) Legacy Imaging Surveys (LIS, \citealp{Dey_2019}) data (taken on 2019/08/08), all markers are the same across both images. Although we see clear spiral arms in the LIS image, the SN position is offset from the host.}
    \label{fig:host_aqi}
\end{figure}

SN~2024abvb is spatially offset from its probable host galaxy (host $z=0.039$), therefore host extinction is likely negligible.  We adopt the line‐of‐sight extinction measurement at the SN position, $A_{V} = 0.507$~mag, corresponding to $E(B-V)_{\mathrm{MW}} = 0.164$~mag  \citep{2011ApJ...737..103S}, as our total extinction. Figure \ref{fig:host_aqi} shows a composite Las Cumbres Observatory (LCO) $gri$ image of the SN field $\sim$6 days post explosion together with that of the Legacy Imaging Survey (DR10) showing the host. Relevant SN and host information can be found in Table \ref{tab:object_info}.

The structure of this paper is as follows.  Section~\ref{sec:data} describes the discovery, follow‐up strategy and data reduction procedures for both photometry and spectroscopy.  In Section~\ref{sec:analysis} we present the detailed analysis of the light curves and spectral series. In Section~\ref{sec:host} we discuss the host of SN~2024abvb and provide an analysis of its environment.  Section~\ref{sec:discussion} we combine our findings and discuss the implications for the progenitor system and circumstellar environment of SN~2024abvb. Finally, in Section~\ref{sec:conclusion} we summarise the paper and make conclusions.

\section{Data}
\label{sec:data}

\subsection{Photometry}
\label{sec:photometry}

Our photometric monitoring of SN~2024abvb utilises a heterogeneous suite of facilities and filter systems to achieve broad wavelength coverage (see Table \ref{tab:full_phot}).  Optical imaging in the Johnson–Cousins $B$- and $V$-bands and an Sloan Digital Sky Survey (SDSS)-like $r$-band was obtained with the Andalucia Faint Object Spectrograph and Camera (ALFOSC) on the Nordic Optical Telescope (NOT; \citealp{2010ASSP...14..211D}).  Complementary $V$-band frames were secured using the ESO Faint Object Spectrograph and Camera (EFOSC2) mounted on the New Technology Telescope (NTT; \citealp{1984Msngr..38....9B}).  Additional optical data in the SDSS-like $g$, $r$, $i$ and $z$ filters were acquired with the 1m network of telescopes of the Las Cumbres Observatory (LCO; \citealp{2013PASP..125.1031B}), utilising the Sinistro cameras.
Since instruments with very different passbands were used for the follow-up of SN~2024abvb we checked the uncertainties of passband corrections (e.g., \citealp{2002AJ....124.2100S,2004MNRAS.355..178P}) to standardise photometry to a common system. We used the S3 package \citep{2018MNRAS.475.1046I} and found that such correction is an order of magnitude lower than the telescope photometric uncertainties. Imaging was taken with the Rapid Eye Mount (REM) telescope \citep{2002ASPC..277..449C}. The Near–infrared (NIR) imaging in the $J$-, $H$- and $K$-bands were taken with the REMIR instrument. Optical $V$-, $R$- and $I$- frames were taken with the ROS2 instrument. Data from the Gravitational-wave Optical Transient Observer (\citealp[GOTO; ][]{2022MNRAS.511.2405S}) were obtained as part of regular all-sky surveying. Image reduction and calibration was performed in real-time using the GOTO transient pipeline. Forced aperture photometry at the position of SN~2024abvb was performed to recover the final GOTO light curve.

All CCD data were processed within the IRAF\footnote{The Image Reduction and Analysis Facility (IRAF) is distributed by the National Optical Astronomy Observatory, operated by AURA under cooperative agreement with the National Science Foundation.} environment.  Standard reduction steps including overscan correction, bias subtraction, flat‐fielding and trimming were applied, and SN photometry was extracted via point‐spread function fitting on the final reduced frames.

Ultraviolet observations were obtained with the Neil Gehrels \textit{Swift} Observatory’s Ultraviolet/Optical Telescope (UVOT; PI: Farias) through the $uvw2$, $uvm2$, $uvw1$, $u$, $b$ and $v$ filters.  These data were calibrated to the Vega system and reprocessed independently using a custom pipeline based on the HEASARC software suite.

To further supplement our proprietary measurements, we incorporated archival photometry from several time‐domain surveys.  These include the Zwicky Transient Facility (ZTF; $g$, $r$ filters; \citealp{2019PASP..131a8002B}), Pan‐STARRS1 ($w$ filter; \citealp{chambers2019panstarrs1surveys}), the BlackGEM array ($q$ filter; \citealp{2024PASP..136k5003G}), the Gravitational‐wave Optical Transient Observer (GOTO; $L$ filter; \citealp{Steeghs_2022,dyer2024gravitationalwaveopticaltransientobserver}) and the ATLAS survey (cyan and orange filters; \citealp{2012DPS....4421012J}).  Non‐public Pan‐STARRS1 measurements were obtained under prior collaboration agreements. Final magnitude measurements from these surveys were retrieved using their data pipelines \citep{2006amos.confE..50M,Magnier_2020}.

\subsection{Spectroscopy}
\label{sec:Spectroscopy}

Nine epochs of spectroscopic data were acquired with the EFOSC2 instrument mounted on the ESO New Technology Telescope (NTT) as part of the ePESSTO+ collaboration.  We obtained low‐resolution spectra using grisms 11 and 16, while medium‐resolution observations were secured with grism 18 (see Table~\ref{tab:spectra_obs} for the full observation log, including wavelength coverage and resolving powers).  The raw frames were processed and calibrated using the Public ESO Spectroscopic Survey of Transient Objects (PESSTO) pipeline\footnote{https://github.com/svalenti/pessto} \citep{2015A&A...579A..40S}, which performs bias subtraction, flat‐field correction, wavelength calibration and flux calibration in a uniform and reproducible manner. All ePESSTO+ spectra will be available through the ESO Science Archive Facility as standard phase 3 ESO products. All spectra are available on WISeREP\footnote{https://www.wiserep.org} \citep{2012PASP..124..668Y}.

\begin{figure}
    \includegraphics[width=\columnwidth]{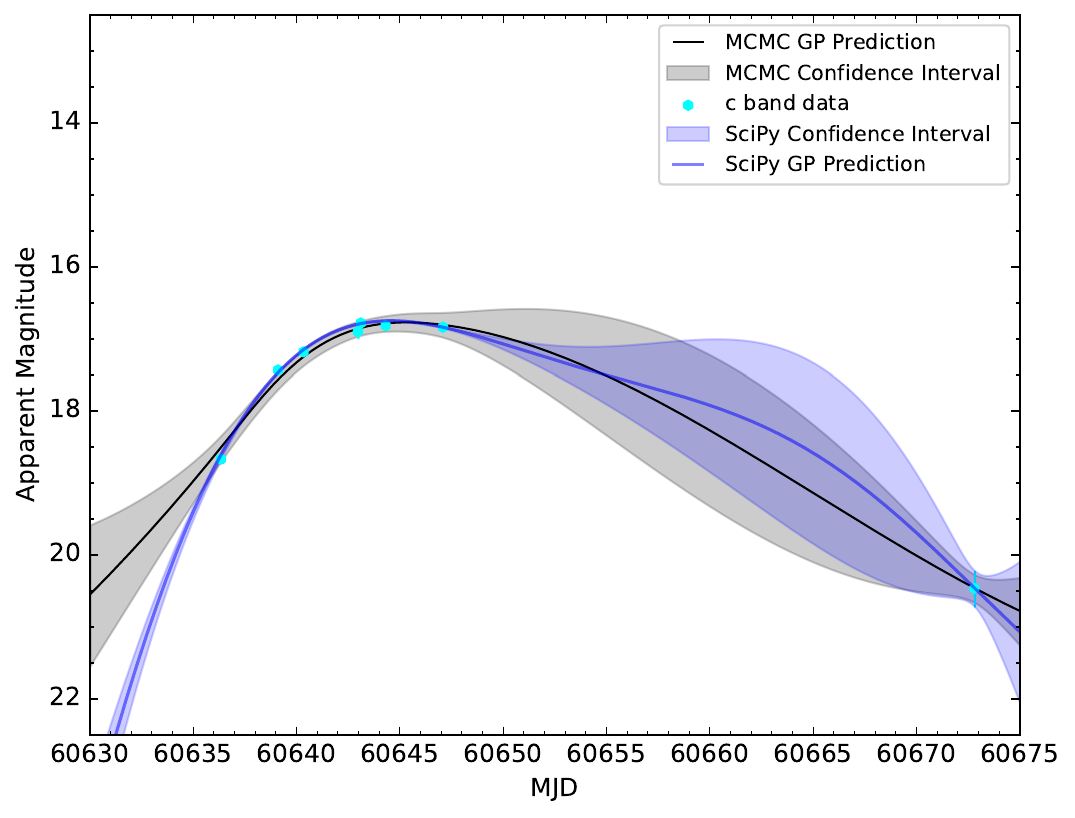}
    \caption{The \textsc{SciPy} and \textsc{emcee} optimised fit for the $c$-band data}
    \label{fig:c_gpfit}
\end{figure}

\begin{table}
\centering
\caption{Key information for the SN and host. Time of last non-detection is 60634.5 MJD in the $o$-band. We adopt the line‐of‐sight extinction measurement $E(B-V) = 0.164$, at the SN position \citep{2011ApJ...737..103S}. No evident reddening is observed in the host (e.g. no narrow absorption lines). Hence, the MW reddening is the total.}
\label{tab:object_info}
\begin{tabular}{lr}
\hline\hline
Time of First Detection (MJD) & 60636.62 \\
Estimated Time of Explosion (MJD) & 60635.45 $\pm$ 1.1 \\
Estimated Time of Maximum (MJD) & 60645.13 $\pm$ 0.4\\
RA (J2000) & 01:10:57.53 \\
Dec (J2000) & -05:44:07.91 \\
Redshift & 0.039 \\
$E(B - V)_{\mathrm{MW}}$~[mag] & 0.164 \\
$m_{V}^{\mathrm{peak}}$~(mag) & 16.70 \\
$M_{V}^{\mathrm{peak}}$~(mag) & $-19.43$ \\
$t_{\mathrm{rise,g/C}}$ (days) & 10.1 \\
$t_{1/2\mathrm{,decline,g/C}}$ (days) & 7.6 \\
Separation from host (arcmin) & 0.464 \\ 
Separation from host (kpc) & 21.5 \\ \hline
\end{tabular}
\end{table}

\section{Characterising SN~2024abvb in the context of interacting supernovae}
\label{sec:analysis}

In this section, we compare photometry and spectra of SN~2024abvb to other SNe, chosen based on the fact that are well sampled and studied SNe. The chosen comparison Type Icn are reported above in Section \ref{sec:intro} while the others are: SN~2010al (Type Ibn, \citealp{2015MNRAS.449.1921P}); SN~2023emq (Type Ibn/Icn, \citealp{2023ApJ...959L..10P}) and iPTF14aki (Type Ibn, \citealp{2017ApJ...836..158H}).

\begin{figure*}
\includegraphics{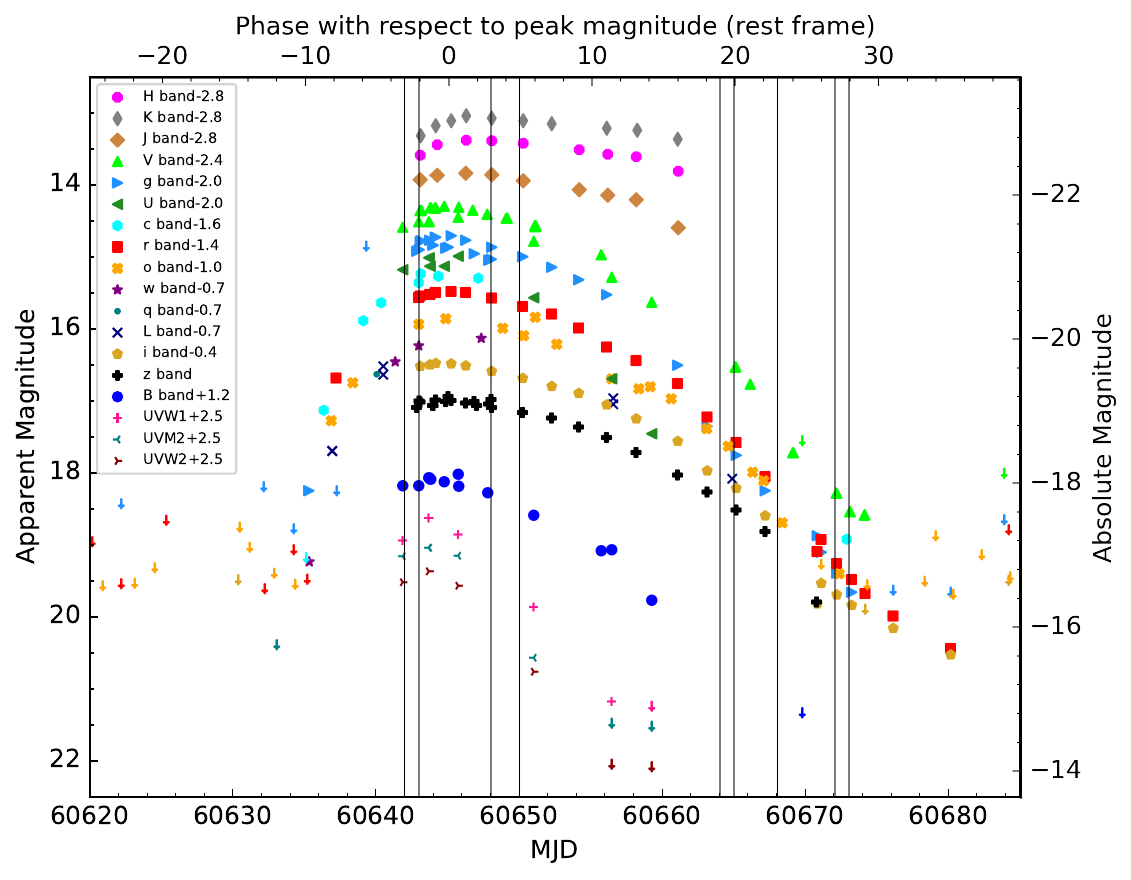}
    \caption{Light curves of SN~2024abvb from AFOSC, ATLAS, Pan-STARRS, LCO, NTT, Swift, NOT, ZTF, REM, TTT, Moravian, BlackGem and GOTO for all available bands from the UV-NIR. The phase is given relative to the $V$-band maximum. The legend indicates which shape and colour is associated with specific bands. Non-detections are shown with downward arrows. The rise times and decline rates for each band can be seen in Table \ref{tab:rise_times}. The vertical black lines indicate the epochs when spectra were taken. The $L$- and $q$-bands refer to GOTO and BlackGem bands, respectively. The figure is not extinction corrected. $griz$ are AB magnitudes, as are those of the surveys ATLAS, PS1 and GOTO. $UBVJHK$ are Vega magnitudes.}
    \label{fig:photometry}
\end{figure*}

\begin{figure}
	\includegraphics[width=\columnwidth]{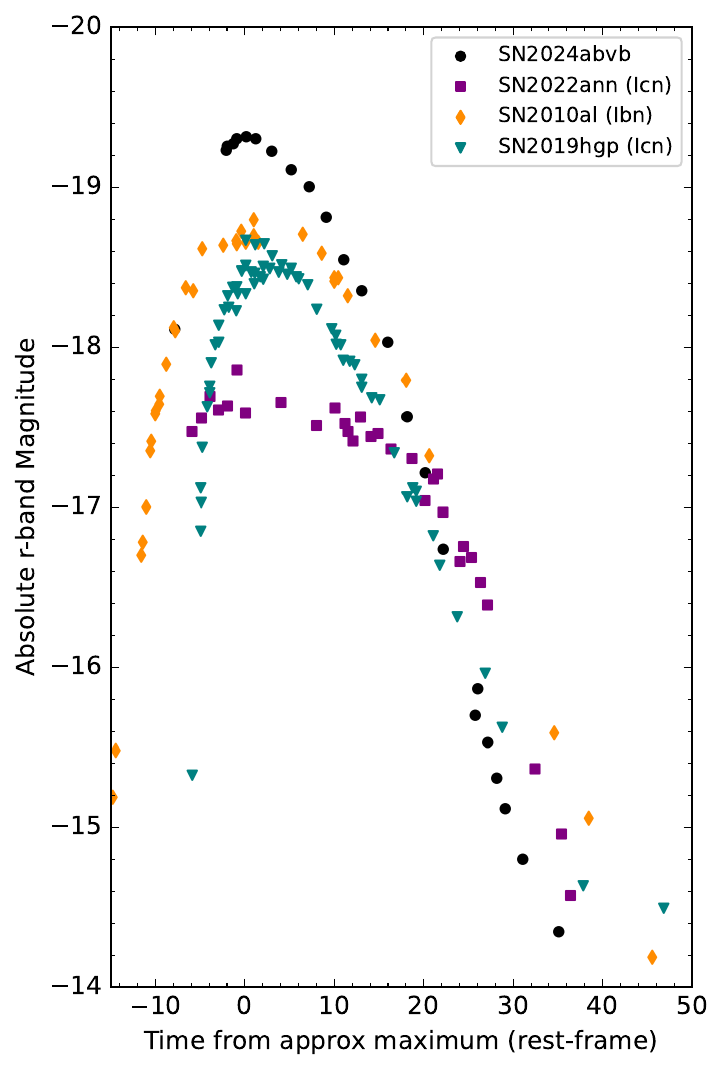}
    \caption{Comparison photometry in the $r$-band (absolute magnitude) for SN~2024abvb, the Type~Ibn (SN~2010al; \citealp{2015MNRAS.449.1921P}) and two other Type~Icn (SNe~2022ann and 2019hgp; \citealp{Davis_2023}).}
    \label{fig:photometry_comp}
\end{figure}

\begin{figure}
	\includegraphics[width=\columnwidth]{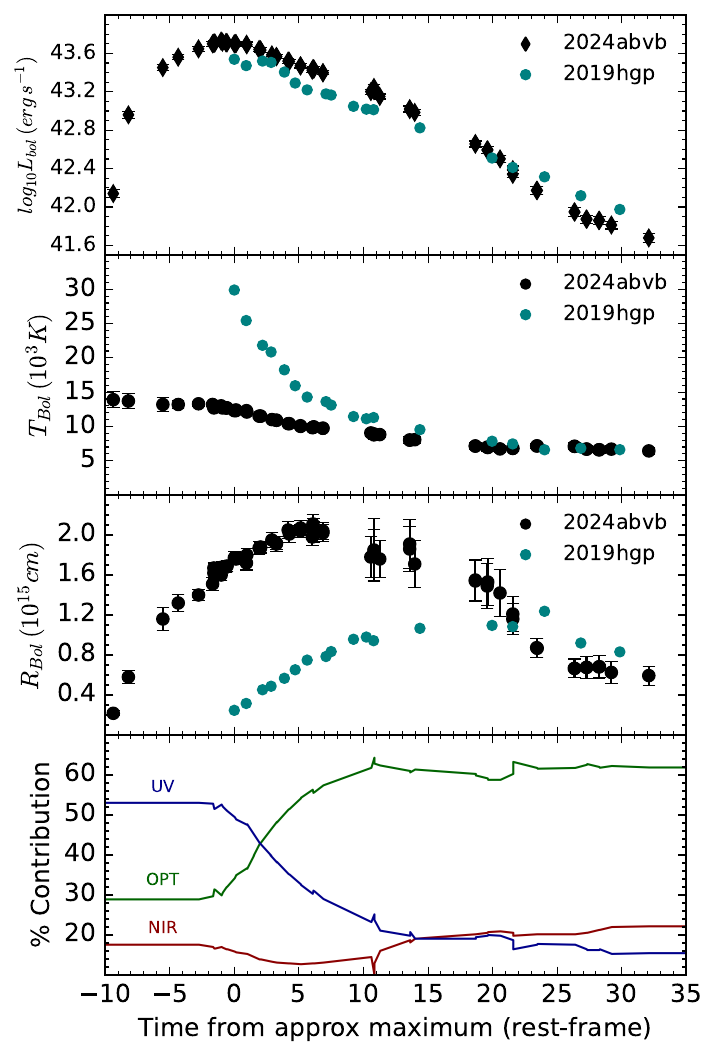}
    \caption{The top panel shows the bolometric light curve. The second panel shows the estimated temperature derived from the black-body fits. The third panel shows the photospheric radius. The bottom panel shows the contribution of each part of the electromagnetic spectrum over time (UV: $uvw2$, $uvm2$, $uvw1$. Optical: $u$, $B$, $g$, $V$, $c$, $r$, $o$, $i$, $z$. NIR: $J$, $H$, $K$). Each panel, besides the last, shows a comparison between SNe~2024abvb and 2019hgp \citep{2022Natur.601..201G}. SN~2019hgp is the closest photometric analogue, therefore we use this for comparison.}
    \label{fig:bol}
\end{figure}

\begin{figure*}
    \includegraphics{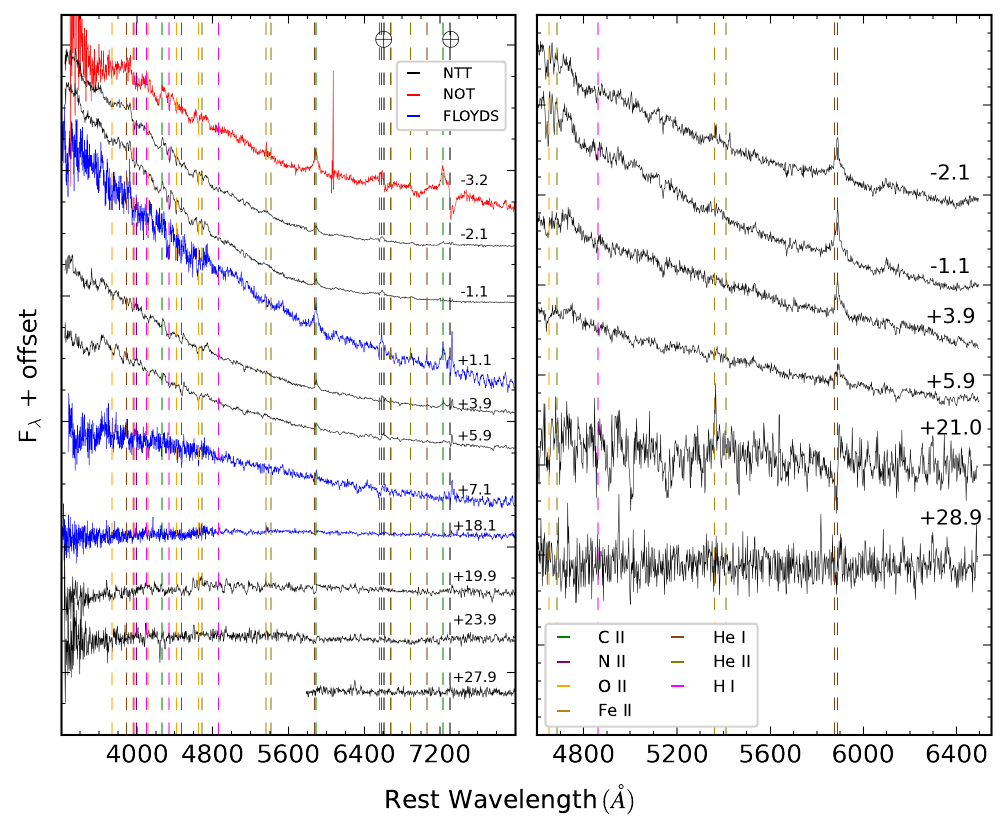}
    \caption{\textbf{Left panel:} All low-resolution spectra (see Table \ref{tab:spectra_obs}), corrected for redshift and reddening, for SN~2024abvb taken between $-$3.23 days and 27.90 days relative to $V$-band maximum brightness. The C~{\sc ii}, H~{\sc i}, He~{\sc i}, He~{\sc ii}, O~{\sc ii}, N~{\sc ii} and Fe~{\sc ii} lines are marked. We see some telluric contamination; however, it leaves the spectral lines mostly unaffected. The cross in this figure shows the telluric lines. \textbf{Right panel:} All medium resolution spectra (see Table \ref{tab:spectra_obs}), corrected for redshift and reddening, for SN~2024abvb taken between $-$2.05 days and 28.90 days relative to $V$-band maximum brightness. In all instances, multi-component decomposition was fully resolved, as were the subsequent P-Cygni profiles.}
    \label{fig:full_spectra}
\end{figure*}

\subsection{Light curve}
\label{sec:LC}

To determine the epoch of maximum light, we first fitted the ATLAS $c$‐band light curve, chosen for its dense temporal sampling from first detection through peak, using a 1D Gaussian process (GP) regression implemented in the \textsc{George} package \citep{2015ITPAM..38..252A} with an exponential‐squared kernel.  Initial kernel hyperparameters were optimised via \textsc{SciPy} minimisation, after which we employed the \textsc{emcee} Markov Chain Monte Carlo (MCMC) sampler \citep{Foreman_Mackey_2013} to sample from the posterior distribution of the GP model parameters.
The MCMC‐refined GP fit closely matches the initial optimisation. For completeness, we also applied an analogous GP fit to the bolometric light curve, retrieving a similar result.  From these analyses, we adopt MJD~60645.13 $\pm$ 0.4  as the time of maximum light (defined as epoch~0).  The explosion epoch was estimated as the midpoint between the first $w$‐band detection and the preceding $o$‐band non‐detection at 0.3~mag deeper, yielding MJD~60635.45~$\pm$~1.1.

Using the MCMC‐optimised GP model, we computed rise times from explosion to maximum light and 20‐day decline rates in each filter.  The rise time, $t_{\rm rise}$, is defined as the interval between the explosion epoch and the GP‐determined peak MJD.  Decline rates ($\Delta m_{20}$) were measured as the difference in magnitude between peak and 20 days post‐peak.  In the $V$-band we find $\Delta m_{20}=2.6\pm0.3$ mag, whereas in the $i$-band $\Delta m_{20}=1.1\pm0.3$ mag. The slower decline in redder bands is not unusual in CCSNe and even more so in interacting SNe, as it suggests reprocessing of high‐energy (X‐ray/UV) photons into optical wavelengths \citep[e.g.][]{1994ApJ...420..268C}.

Figure~\ref{fig:photometry} displays the UV–NIR light curves of SN~2024abvb.  The $g$‐band rise time of $10.1\pm0.3$ days is notably longer than the $\lesssim5.2$ day values typical of other Type~Icn events (Table~\ref{tab:rise_times}, \citealp{Pellegrino_2022, Perley_2022, 2023A&A...673A..27N}).  Similarly, the ATLAS $o$‐band rise time of $10.5\pm0.4$ days slightly exceeds the $\sim8$ days measured for SN~2022ann \citep{Davis_2023}.  Across all bands, we observe approximately linear declines in magnitude, with the steepest fades in the UV and blue optical filters and the gentlest in the NIR.

In Figure~\ref{fig:photometry_comp} we compare the $r$‐band light curve of SN~2024abvb with those of well-sampled interacting SNe in the same time frame of SN~2024abvb: SNe~2019hgp (Type~Icn), 2022ann (Type~Icn) and 2010al (Type~Ibn).  SN~2024abvb reaches peak $r$‐band light ($M_r\approx-19.4$) at $12.1\pm0.3$ days and shows a $\Delta m_{20}(r)=1.6\pm0.3$ mag.  For reference, SN~2019hgp exhibits a similar $\Delta m_{20}(r)=1.6\pm0.2$ mag but peaks at a lower luminosity ($M_r\approx-18.6$), whereas the prototypical Type~Ibn SN~2010al fades more slowly, with $\Delta m_{20}(r)=1.4\pm0.2$ mag.  SN~2022ann displays a two‐stage decline: an initial slow phase ($\sim0.2$ mag per 20 days) followed by a rapid drop ($\sim2.0$ mag per 20 days), indicative of interaction with a denser CSM. SN~2019hgp has a low-mass CSM environment (0.2~$M_{\odot}$, \citealp{2022Natur.601..201G}). Assuming SN~2024abvb light curve is dominated by the CSM interaction, the similarity of its decline to that of SN~2019hgp could suggest a comparable low‐mass CSM environment for SN~2024abvb and progenitor scenario. However, SN~2024abvb also has a higher peak luminosity compared to SN~2019hgp ($M_r\approx-19.2$ mag), consistent with a somewhat more efficient radiative conversion \citep{2017hsn..book..875C}.

\begin{table}
\caption{Comparison of rise-times and half decline rates in the $g$-band for SN~2024abvb and all Type~Icn SNe where they were measurable. We see that the $g$-band rise time for SN~2024abvb is much higher than for other Type~Icn SNe, while the half decline rate is comparable.}
\label{tab:rise_times}
\centering
\begin{tabular}{lcc}
\hline
SN & Rise Time & Half decline rate\\
 & (days) & (days)\\\hline
2024abvb & 10.1~$\pm$~0.3 & 7.6~$\pm$~0.3\\
2019jc & 3.2~$\pm$~0.1 \textsuperscript{\textit{a}} & 3.1~$\pm$~0.1 \textsuperscript{\textit{a}}\\
2019hgp & 5.2~$\pm$~0.2 \textsuperscript{\textit{a}} & 8.0~$\pm$~0.2 \textsuperscript{\textit{a}}\\
2021csp & 1.8$-$4.0 \textsuperscript{\textit{b}} & 9.1~$\pm$~0.8 \textsuperscript{\textit{b}}\\
2021ckj & 3.3~$\pm$~0.2 \textsuperscript{\textit{c}} & 4.7~$\pm$~0.2 \textsuperscript{\textit{c}}\\ \hline 
\multicolumn{3}{l}{\textit{a} \cite{Pellegrino_2022}} \\
\multicolumn{3}{l}{\textit{b} \cite{Perley_2022}} \\
\multicolumn{3}{l}{\textit{c} \cite{2023A&A...673A..27N}} \\
\end{tabular}
\end{table}

\subsection{Bolometric light curve}
\label{sec:BLC}

Bolometric luminosities were derived by first converting the extinction‐corrected broadband magnitudes (Section~\ref{sec:LC}) into monochromatic fluxes at each filter’s effective wavelength.  A spectral energy distribution (SED) was then assembled over the observed wavelength range, and the integrated flux, $F_{\rm bol}$, was computed under the assumption of negligible contribution beyond the integration limits.  Luminosities were obtained via 
\[
L_{\rm bol} \;=\; 4\pi D^{2} \, F_{\rm bol},
\]
where $D$ is the previously determined distance to SN~2024abvb (Section~\ref{sec:intro}).

Pseudo‐bolometric points were initially calculated for epochs with concurrent coverage in at least four optical bands; for epochs with fewer than four filters, missing fluxes were estimated by low‐order ($n\leq3$) polynomial interpolation of adjacent light‐curve data.  When interpolation was not possible, magnitudes were extrapolated by assuming constant colours between the nearest epochs.

The resulting bolometric light curve (Figure~\ref{fig:bol}) reaches a peak luminosity of $\log(L_{\rm bol}/{\rm erg\,s^{-1}})=43.7$ in approximately 10 days, followed by a nearly linear decline over the subsequent $\sim30$ days.  In the second panel of Figure~\ref{fig:bol}, we show the black‐body temperature, $T_{\rm BB}$, inferred from SED fits. $T_{\rm BB}$ attains a maximum of roughly 14000~K, remains on a quasi‐plateau, then decreases rapidly before settling at around 6400~K around 20 days post‐peak.  These temperatures are substantially lower than those reported for SNe~2022ann and 2019hgp, which peaked near 25000~K and 30000~K, respectively.

The third panel of Figure~\ref{fig:bol} presents the photospheric radius 

\[
R_{\rm ph} \;=\; \sqrt{\dfrac{L_{\rm bol}}{4\pi\sigma T_{\rm eff}^4}},
\]

\noindent where $\sigma$ is the Stefan–Boltzmann constant.  $R_{\rm ph}$ increases to a maximum of $\sim2.1\times10^{15}$~cm at $\sim6$~days after peak, then declines slowly and plateaus at $\sim0.65\times10^{15}$~cm by $\sim25$~days post‐peak. Overall, the estimated radius of SN~2024abvb is larger than that showcased by SN~2019hgp, but at a size not unusual for SNe \citep[e.g.][]{Kasen_2009, 2022A&A...658A.130D,2024ApJ...973...14C}.

The bottom panel of Figure~\ref{fig:bol} shows the fractional flux contributions from the ultraviolet (UV), optical and near‐infrared (NIR) regimes.  Initially, the UV accounts for $\sim53\%$ of $F_{\rm bol}$ and the optical for $\sim29\%$, with the NIR contributing $\sim17\%$.  Over the first two days post‐maximum, the optical fraction rises to match the UV at $\sim43\%$, consistent with high‐energy photon reprocessing by circumstellar material (CSM).  The NIR fraction decreases to $\sim12\%$ in the early decline, then increases slightly as the optical component plateaus, ultimately stabilising just above its initial level.  Within our measurement uncertainties, the NIR contribution remains effectively constant, implying that any dust formation either is minimal or occurs at epochs later than our final NIR observations \citep{2008MNRAS.389..141M}.

\subsection{Spectral evolution and line measurements}
\label{sec:spec_evo}

Figure~\ref{fig:full_spectra} displays the spectral evolution of SN~2024abvb, spanning from three days before maximum light to approximately one month thereafter.  Prominent transitions of C~{\sc ii} and O~{\sc ii} are marked alongside the tentative detections of H~{\sc i} , He~{\sc i} and He~{\sc ii}. The continua are consistent with the photospheric temperatures derived from our black‐body fits.

To quantify the kinematic and flux properties of individual lines, we fitted Gaussian profiles using the \textsc{Lmfit} package\footnote{\textsc{Lmfit} is a non‐linear least‐squares minimisation and curve‐fitting library for Python.} \citep{newville_2014_11813}.  For the C~{\sc ii} $\lambda5890$ feature, a composite model comprising a central Lorentzian (narrow emission) plus a broader Gaussian component was employed to capture both the line core and extended wings.

In the four low-resolution spectra obtained near peak brightness, the continuum is blue and several narrow P-Cygni profiles are evident, with C~{\sc ii} $\lambda5890$ dominating in emission.  At early epochs (0–4~days relative to peak), this line exhibits a full width at half maximum (FWHM) of $\sim41$~\AA, corresponding to a velocity of $\sim2100\,$km~s$^{-1}$, and an integrated flux of $1.0\times10^{-16}$~erg~s$^{-1}$~cm$^{-2}$.  By 6~days post‐peak, the FWHM decreases to $\sim32$~\AA\/ (velocity $\sim1600\,$km~s$^{-1}$) with a flux of $8.8\times10^{-17}$~erg~s$^{-1}$~cm$^{-2}$.  Beyond 20~days, the continuum flattens, likely due to metal‐line blanketing below 4700~\AA, and C~{\sc ii} $\lambda5890$ evolves into a P-Cygni profile with FWHM $\sim28$~\AA\/ (velocity $\sim1400\,$km~s$^{-1}$) and flux $2.5\times10^{-17}$~erg~s$^{-1}$~cm$^{-2}$.

\begin{figure}
    \includegraphics[width=\columnwidth]{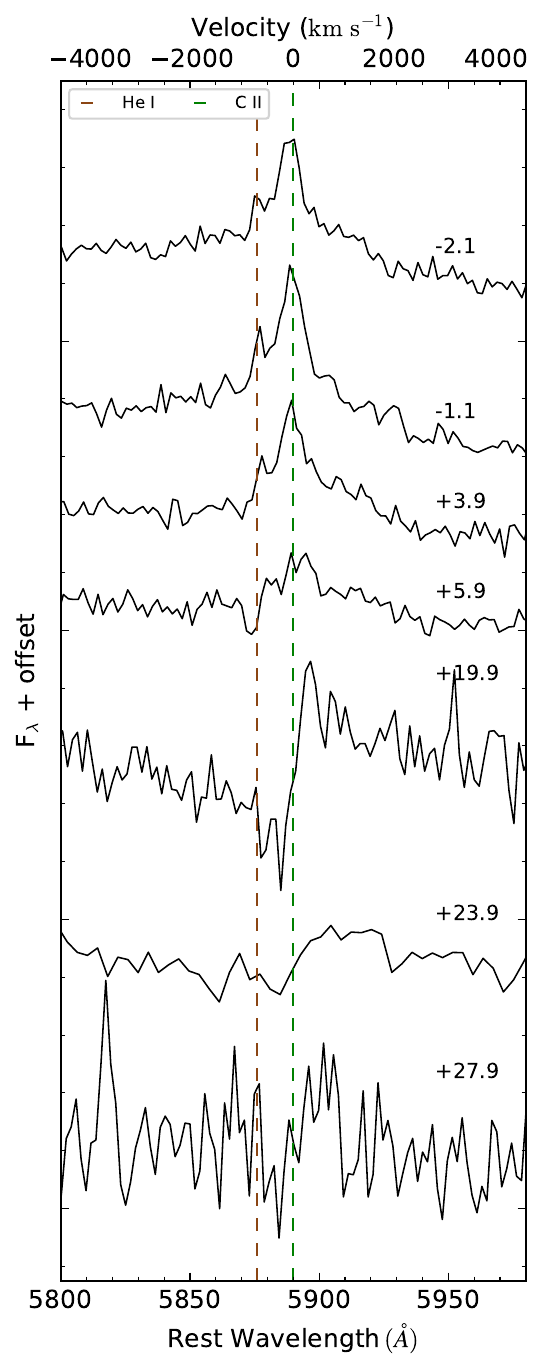}
    \caption{The same spectra are shown here as in Figure \ref{fig:full_spectra}; however, this is a zoom-in on the most prominent spectral feature, C~{\sc ii} $\lambda5890$~\AA, with medium resolution spectra (see Table \ref{tab:spectra_obs}).}
    \label{fig:full_spectra_cii}
\end{figure}

Medium‐resolution spectra (grism~\#18 at the NTT+EFOSC2; Table~\ref{tab:spectra_obs}) enable a detailed view of the 4600–6500~\AA\/ region (Figure~\ref{fig:full_spectra_cii}).  Here, C~{\sc ii} $\lambda5890$ has FWHM $32.7$~\AA\/ (velocity $\sim1600\,$km~s$^{-1}$) and flux $1.5\times10^{-16}$~erg~s$^{-1}$~cm$^{-2}$ at peak, evolving to FWHM $29.1$~\AA\/ (velocity $\sim1400\,$km~s$^{-1}$) and flux $7.6\times10^{-17}$~erg~s$^{-1}$~cm$^{-2}$ by +6~days. And at $\sim30$~days FWHM $16.8$~\AA\/ (velocity $\sim800\,$km~s$^{-1}$) and flux $2.88\times10^{-17}$~erg~s$^{-1}$~cm$^{-2}$. All FWHM measurements are above our spectra resolution (see Table \ref{tab:spectra_obs}).
A weaker feature at $\sim5876$~\AA, present until +4~days, is consistent with being He~{\sc i} $\lambda5876$. 
The red wing of C~{\sc ii} $\lambda5890$ exhibits a broad shoulder which, as mentioned above, can be due to metal‐line blanketing in the blue part or suggestive of asymmetric CSM.

O~{\sc ii} $\lambda4651$ is detected in low‐resolution spectra prior to +20~days, with an early FWHM of $9.3$~\AA\/ (flux $1.2\times10^{-16}$~erg~s$^{-1}$~cm$^{-2}$) broadening to $26.0$~\AA\/ (flux $9.4\times10^{-17}$~erg~s$^{-1}$~cm$^{-2}$) by +6~days; it is absent at later epochs.  In contrast, C~{\sc ii} $\lambda5890$ persists, switching to a P-Cygni morphology in both low‐ and medium‐resolution spectra after +20~days (FWHM $19.4$–$27.8$~\AA, flux $\sim4.0\times10^{-17}$~erg~s$^{-1}$~cm$^{-2}$).

Additional features, including H~{\sc i}~$\lambda4861$, O~{\sc ii}~$\lambda4651$ and Fe~{\sc ii}~$\lambda5363$, are discernible in medium‐resolution data, with clear P-Cygni profiles persisting at late times.

\begin{figure*}
    \includegraphics{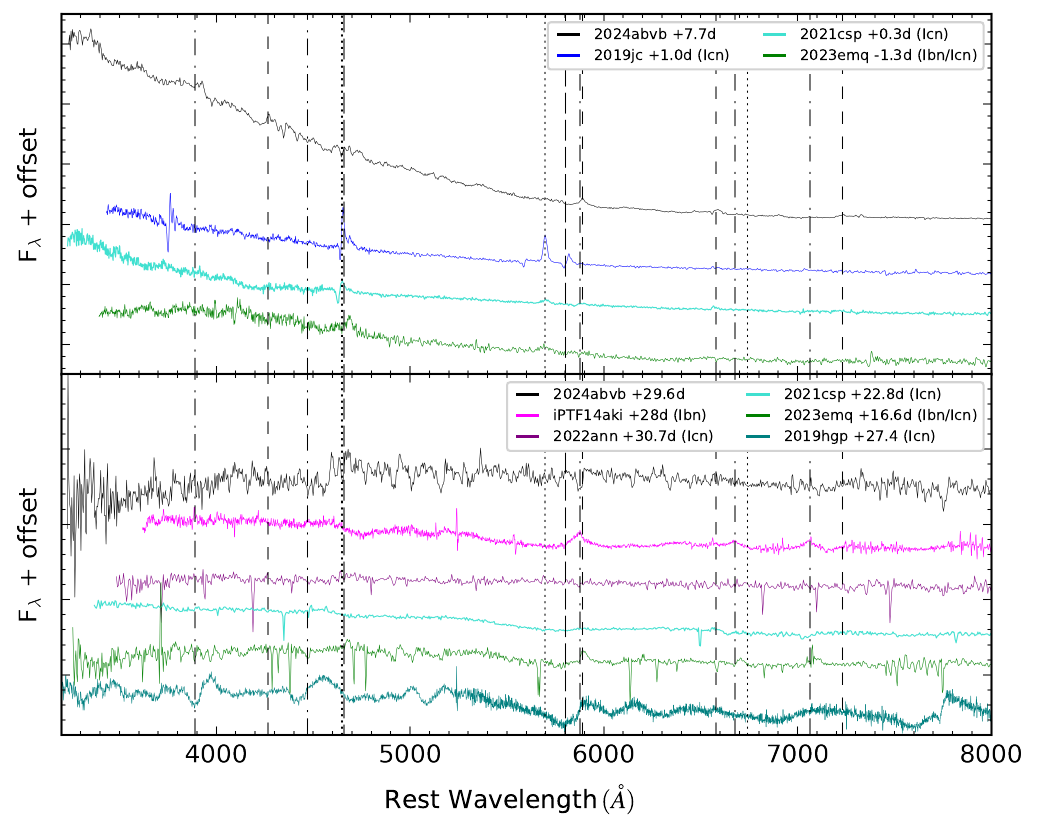}
    \caption{SN~2024abvb spectra compared to other Type Ibn, Type Icn and transitional Type Ibn/Icn SNe. Both panel have dashed lines to depict different elements: C~{\sc ii}  \hdashrule[0.5ex]{0.75cm}{0.75pt}{1mm}; C~{\sc iii} \hdashrule[0.5ex]{0.75cm}{0.75pt}{1pt}; C~{\sc iv} \hdashrule[0.5ex]{1cm}{0.75pt}{3mm 3pt};  He~{\sc i} \hdashrule[0.5ex]{1cm}{0.75pt}{2mm 2pt 0.25mm 1pt}; \textbf{Top panel:} Early time ($\sim$~8 days) spectra comparison for SN~2024abvb, SN~2018fmt, SN~2019jc, SN~2021csp and SN~2023emq. The specific colours and phase times for each SN are shown in the legend. The phase is with respect to the explosion time as we did not have access to photometric data for the comparison SNe. \textbf{Bottom panel:} Late time ($\sim$~29 days) spectra comparison for SN~2024abvb, SN~2021csp, iPTF14aki, SN~2022ann, SN~2023emq and SN~2019hgp.}
    \label{fig:comp_combined_2}
\end{figure*}

\subsection{Spectral comparison with other Type~Ibn, Type~Icn and transitional events}
\label{sec:spec_comp}

Figure~\ref{fig:comp_combined_2} displays the EFOSC2 spectra of SN~2024abvb at epochs near maximum light and approximately one month later, compared with those of SNe~2018fmt, 2019jc, 2021csp, 2023emq \citep{2023ApJ...959L..10P}, iPTF14aki and 2022ann. Comparison SNe were chosen based on having spectra covering from peak epoch to roughly a month later in their evolution. The sample includes well studied Type~Ibn events (SNe~2018fmt and iPTF14aki), canonical Type~Icn events (SNe~2019jc and 2021csp) and recognised transitional Type~Ibn/Icn objects (SN~2023emq), chosen to assess whether SN~2024abvb exhibits intermediate characteristics.

At early times, SN~2024abvb exhibits a notably blue continuum, closely matching the continua of SNe~2021csp and 2023emq, remaining bluer than those of the other Type~Ibn and Type~Icn comparisons.  The persistence of this blue slope, also evident in the minimal $B$‐band decline of $\Delta m(B)\approx0.16$~mag between the first and third spectra, suggests more efficient conversion of high‐energy (X‐ray/UV) photons into optical wavelengths than in typical Type~Icn events. We note that this is at odds with what is observed in the temperature evolution.
In contrast, SNe~2019jc and 2021csp display a pseudo‐continuum dominated by blended emission lines \citep[e.g.][]{2022Natur.601..201G}, which is not apparent in SN~2024abvb.  Although Figure~\ref{fig:comp_combined_2} hints at a weak pseudo-continuum between 4600 and 6000~\AA\/ in SN~2024abvb, this feature remains ambiguous.

In the bottom panels of Figure~\ref{fig:comp_combined_2}, SN~2019hgp, the closest photometric analogue to SN~2024abvb, exhibits broader and faster line profiles and displays intermediate‐mass element features such as Mg~{\sc i}, which are absent in SN~2024abvb.  Moreover, oxygen emission lines in SN~2019hgp are considerably stronger, and its late‐time spectrum develops the characteristic iron pseudo‐continuum common to many Type~Icn (and interacting) events. SN~2021csp, similarly to SN~2019hgp, shows an Fe pseudo-continuum but otherwise shows very few emission or absorption features. SN~2022ann shows a much flatter continuum than the other three Type~Icn presented, but similar to SN~2021csp it shows very few prominent absorption and emission features. This is in stark contrast to SNe~2024abvb and 2019hgp, which both display prominent features in pure emission. 
The one transitional SN Type~Ibn/Icn we present shows an interesting blueward feature, similar to that of SN~2024abvb; however, on further inspection, this is more likely due to an Fe pseudo-continuum as seen in other Type~Icn SNe. Such a continuum seems lacking in SN~2024abvb where we see a sharp decrease at $\sim$4500~\AA, implying this feature is formed through an alternative mechanism.
We observe that iPTF14aki shows a flatter continuum with characteristically broad Helium lines typically seen in late-time Type~Ibn spectra \citep{2017ApJ...836..158H}. This, along with the other features noted in SN~2024abvb, gives strong evidence that SN~2024abvb is inconsistent with a Type~Ibn SN classification.

The close temporal comparison reveals that SN~2019hgp has elevated flux compared to SN~2024abvb, both blueward of 4600~\AA\/ and redward of 6200~\AA\/, whereas between these wavelengths (4600–6200~\AA\/) the fluxes of the two SNe are broadly comparable, with both displaying a flux enhancement near 6000~\AA.  These differences reinforce the classification of SN~2024abvb as a transitional Type~Icn/Ibn object with a relatively low‐mass, carbon‐rich CSM.

\subsection{Expansion velocities of the CSM}
\label{sec:vel}

\begin{figure}
	\includegraphics[width=\columnwidth]{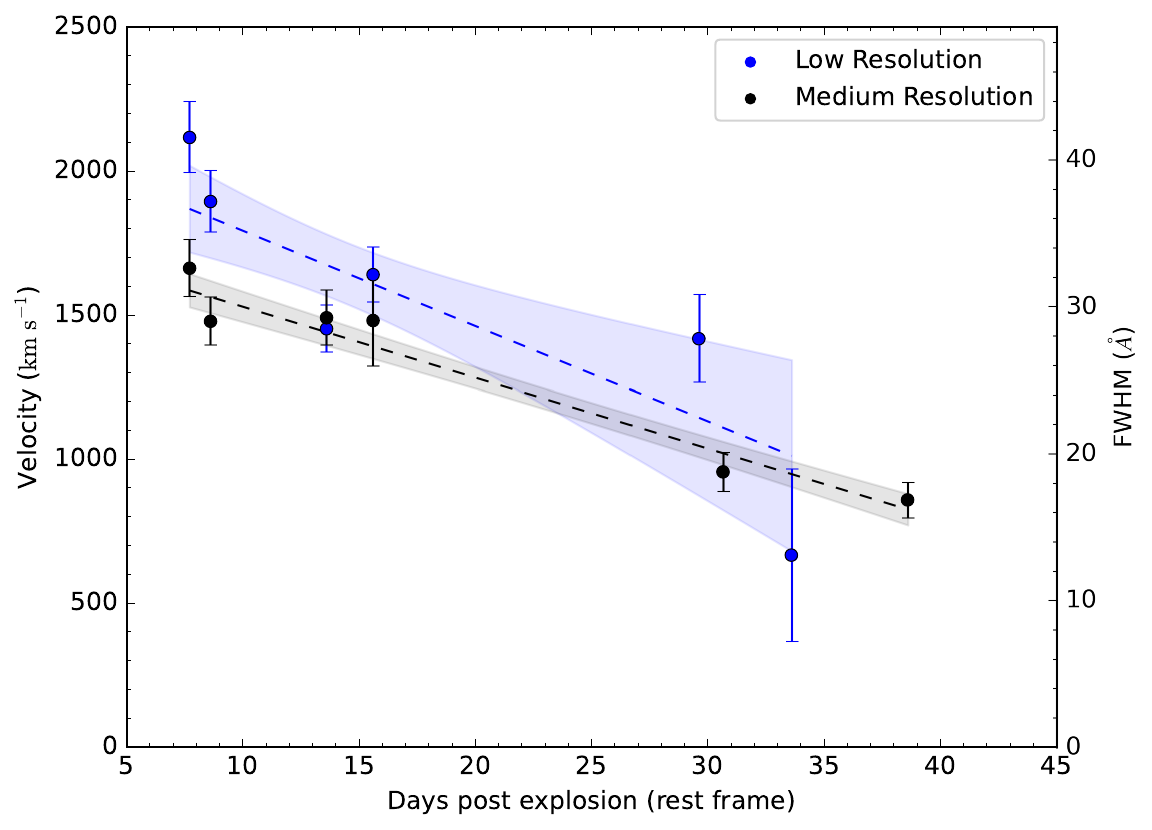}
    \caption{C~{\sc ii} 5890~\AA\/ comparison velocities for low and medium resolution spectra. The first four epochs were measured using the emission component FWHM, while the remaining two were measured from P-Cygni minima. The dashed line depicts the best-fit while the shaded regions display the $1\sigma$ confidence intervals.}
    \label{fig:cii_5890_vel}
\end{figure}

We derived the expansion velocity of the carbon‐rich CSM from C~{\sc ii} $\lambda5890$.  For epochs up to $\sim$10~days after maximum light, we measured the full width at half maximum (FWHM) of the Gaussian fits, while at later times we determined velocities from the minima of the P-Cygni absorption troughs.  The resulting velocity evolution is plotted in Figure~\ref{fig:cii_5890_vel}.  Differences between low‐ and medium‐resolution measurements are attributable to instrumental resolving power.

As mentioned above, He~{\sc i $\lambda5876$} is also present and alters the overall profile of the C~{\sc ii} line, as such we use our medium resolution spectra measurements as a more accurate diagnostic.
At early epochs, the C~{\sc ii} line exhibits velocities of $\sim1500\,$km~s$^{-1}$, declining smoothly to $\sim800\,$km~s$^{-1}$ by the final measurement. This smooth decline is observed in both the low- and medium-resolution spectra, with $\Delta V_{20-\mathrm{low}} \sim 660\,\pm\,150\,\mathrm{km\,s^{-1}} $ and $\Delta V_{20-\mathrm{med}} \sim 490\,\pm\,100\,\mathrm{km\,s^{-1}}$, respectively. In both cases, the minimum measured values remain above the instrumental resolution (460~km~s$^{-1}$ for gr18 and 870~km~s$^{-1}$ for gr11), although the final velocity point in the medium-resolution data lies close to this threshold.
The final measurement conflicts with early WC and canonical WO wind speeds of $>1000\,$km~s$^{-1}$ \citep{2019A&A...621A..92S,1971ApJ...164..275S}. However, canonical wind speeds for WN and late type WC9 stars are lower at $v_{\rm wind}>650\,$km~s$^{-1}$ and $\sim450\,$km~s$^{-1}$, respectively \citep{2019A&A...625A..57H,1971ApJ...164..275S}. This suggests that the CSM may have been produced by a discrete eruption, by binary interaction, by a WN star in a binary system or by a WC9 star with emission transitioning from lower density CSM (higher velocity lines) to higher density inner regions (lower velocity lines). In the eruptive scenario, the required mass‐loss rate would be sufficiently high to dominate the circumstellar density, whereas a sustained WR wind alone would be less likely to generate the observed density profile; this interpretation is in line with \citet{Smith_2017}.

Comparable velocities have been reported for other interacting SNe.  For example, SN~2019hgp displayed C~{\sc iii} $\lambda4650$ velocities in excess of $1500\,$km~s$^{-1}$, consistent with WR progenitors but quoted as a lower limit due to spectral resolution \citep{2022Natur.601..201G}.  Conversely, SN~2022ann exhibited expansion speeds of $\sim800\,$km~s$^{-1}$. \citet{Davis_2023} favour a binary scenario given the line driven wind speeds are below early WC and WO wind speeds \citep{2019A&A...621A..92S}. The close agreement between the velocity of SN~2022ann and our late‐time measurement for SN~2024abvb supports a binary progenitor system and is therefore inconsistent with a single early WC or WO star progenitor. This suggests that the progenitor scenario is either a WN star, a late type WC9 star or more complex than a single WR star.

\section{Host analysis}
\label{sec:host}
From optical images, SN~2024abvb appeared to be far from any potential host and possibly hostless, with the nearest galaxy located 0.464~arcmin (21.5~kpc) S/W of the SN (Figure~\ref{fig:host_aqi}). 
To determine whether the S/W galaxy is the host, we obtained a combination of Gr11 and Gr16 spectra. The slit was centered on the nucleus and covered the majority of the host. This translates to a mostly global spectra but was only used for determining the redshift.

\begin{figure}
	\includegraphics[width=\columnwidth]{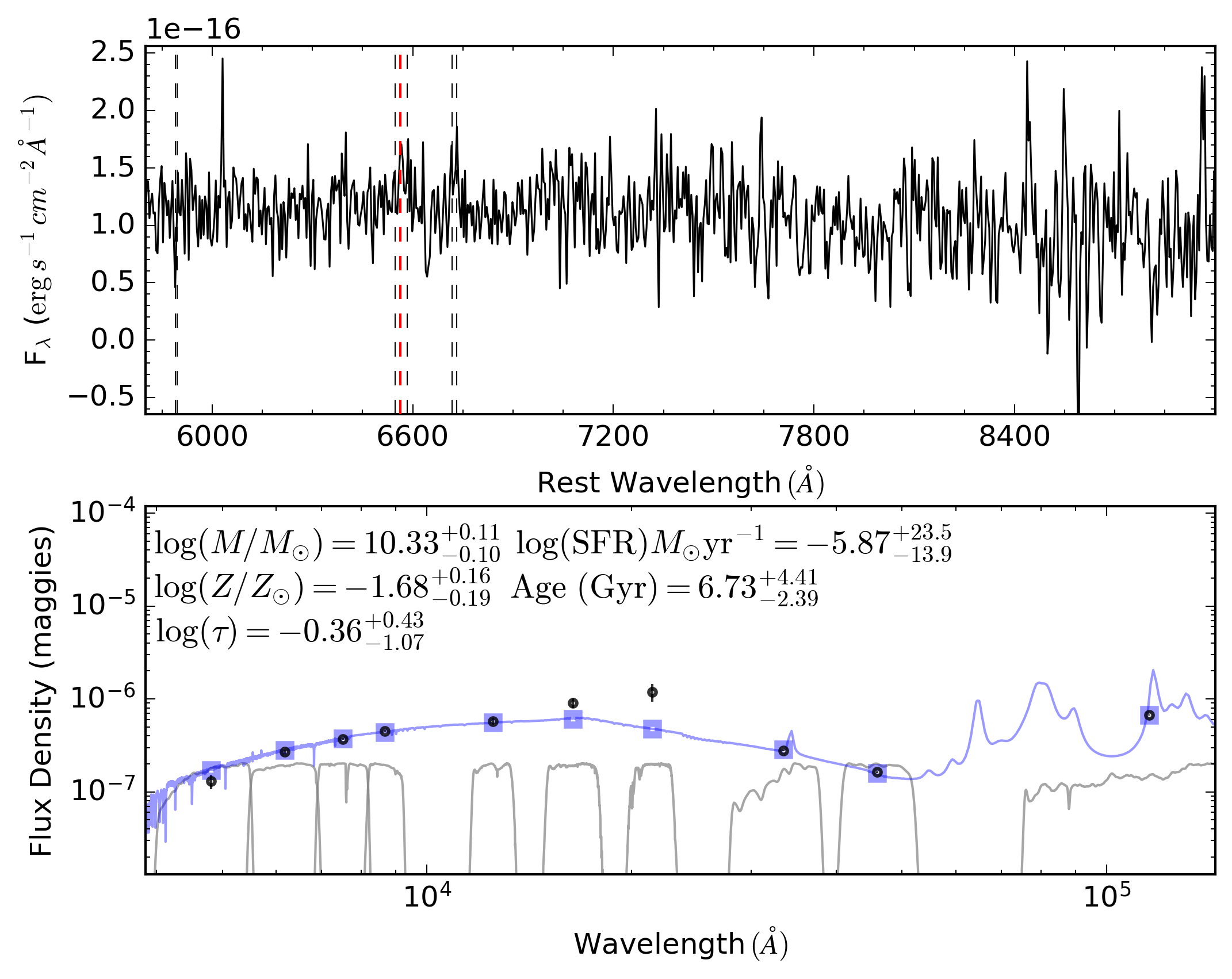}
    \caption{\textbf{Top panel:} Host spectrum with galaxy lines (H$\alpha$ indicated in red, Na~\textsc{i}, N~\textsc{ii} and S~\textsc{ii} are indicated in black) at the assumed redshift of $z=0.039$. \textbf{Bottom panel:} Best fit SED obtained from \textsc{Prospector}. The black points are the LIS $griz$, $W1$,$W2$ and $W3$ and VHS $JHKs$ data points with error bars. The blue boxes and line are the fitted SED and data points. This fit has a reduced $\chi^{2} = 0.025$.}
    \label{fig:host_comb}
\end{figure}

Unfortunately, due to environmental conditions, the Gr11 spectrum had S/N~$\leq$~3 and was therefore unusable. The Gr16 spectrum (upper panel of Figure~\ref{fig:host_comb}) also has low S/N, however, features were visible. A redshift of $z=0.039$, the same value as identified for SN~2024abvb through spectral comparison, appears to confirm this galaxy as the likely host, as H$\alpha$ is detected at the corresponding wavelength.

To retrieve the host galaxy properties, $griz$ and mid-infrared Wide-field Infrared Survey Explorer (WISE) $W1$, $W2$ \& $W3$  magnitudes were taken from the Dark Energy Spectroscopic Instrument (DESI) Legacy Imaging Surveys (LIS, \citealp{Dey_2019}), Data Release 10. The $W4$ was also available but was excluded as that is sensitive to AGN dust emission and we do not implement for such a correction in our simple models, meaning $W4$ would have a poor fit.
We also include $JHKs$ data from the Vista Hemisphere Survey  Data Release 7 (VHS DR7), this survey uses the Visible and Infrared Survey Telescope for Astronomy (VISTA) telescope \citep{2006Msngr.126...41E} using the VISTA infrared camera \citep{2006SPIE.6269E..0XD}.
The LIS data were then input into \textsc{Prospector}\footnote{\textsc{Prospector} is a package to conduct principled inference of stellar population properties from photometric and/or spectroscopic data using flexible models.} \citep{2021ApJS..254...22J} to estimate the host parameters such as stellar mass and star formation rate (SFR). We used the following Prospector parameters: Star formation history is a delayed Tau model; Inital mass function (IMF) is the Chaubrier IMF model \citep{2003PASP..115..763C}; dust type is a MW extinction model \citep{1989ApJ...345..245C}; $\mathrm{log}(Z/Z_{\odot})$ and age are uniform functions; $\mathrm{log}(\tau)$ and $\mathrm{log}(M/M_{\odot})$ are log uniform functions.

The bottom panel of Figure \ref{fig:host_comb} presents a best-fit model for $grizy$, $JHKs$, and $W1$-$W3$ bands given by \textsc{Prospector}. The model provides useful insight into the galaxy's parameters, and has a $\chi^2 = 0.025$ whcih is mainly driven by the optical bands that have lower uncertainties than the near and mid infrared ones. We note that NIR flux is sensitive to intermediate-age stars ($0.5$–$2$~Gyr), also known as Thermally-Pulsing AGB (TP-AGB) stars, and differences of 0.5 dex in JHK from stellar population synthesis models have been widely reported \citep{2009ApJ...699..486C}. This difference is enhanced by the choice of a parametric star formation history (SFH). The Prospector model outputs are $\mathrm{log}(M/M_{\odot})=10.33^{+0.11}_{-0.10}$, $\mathrm{log}(Z/Z_{\odot})=-1.68^{+0.16}_{-0.19}$, $\mathrm{log}(\tau)=-0.36^{+0.43}_{-1.07}$, $\mathrm{Age~(Gyr)}=6.73^{+4.41}_{-2.39}$, $\mathrm{log(SFR)}M_{\odot}\mathrm{yr}^{-1}=-5.87^{+23.5}_{-13.9}$ and presented in the bottom panel of Figure \ref{fig:host_comb}.
The results imply a host galaxy of low mass, slightly above the dwarf galaxy range \citep{Geha_2012}, and well below larger galaxies such as the Milky Way \citep[$\sim10^{12}M_{\odot}$,][]{Watkins_2019}. These galaxies tend to have a lower metallicity and age \citet{2005MNRAS.362...41G}. This is consistent with our results and suggests that the host is likely a young, low-mass, and low-metallicity galaxy. However, because a parametric SFH was adopted, these results should be interpreted as indicative rather than definitive, since a non-parametric SFH would provide a more robust constraint. The peak SFR occurs a $t_{\mathrm{peak}} = \tau \sim 0.28$~Gyr, suggesting that SFR was very high at early time and then decreased, which is consistent with old stellar population. However, such a result mathematically forces the instantaneous SFR at the present epoch to be close to zero ($\sim10^{-6}$). Because delayed-$\tau$ models enforce a monotonic decline in star formation, we caution from over-interpreting the instantaneous SFR derived from this parameterisation. 

The separation of SN~2024abvb from its presumed host also appears atypical compared to other Type~Icn SNe, which are all located within their host galaxies \citep{2022Natur.601..201G, Pellegrino_2022, Davis_2023, Perley_2022, 2023A&A...673A..27N}. However, there are known Type~Ibn~SNe such as, PS1-12sk \citep{Hosseinzadeh_2019}, with a large host separation (28.1~kpc). 
Due to the large separation, we investigate the possibility that SN~2024abvb is hosted by a dwarf satellite galaxy that is gravitationally bound to the host, similarly to the Small Magellanic Cloud (SMC). LIS deep imaging reports depths of $m_{g} = 25.08$, $m_{r} = 25.41$, $m_{i} = 26.86$ and $m_{z} = 23.33$. \citet{dolivadolinsky2025ngc3109satellitesystemsystematic} find that NGC3109, an SMC-like dwarf galaxy has $M_{V}\sim14.9$. This can be roughly converted to $g$ via $g=V+0.60(B-V)-0.12$ where $(B-V)\sim 0.5$ \citep{Sharina_2019}. This gives $M_{g} \sim -15.1$ which when converted to $riz$ gives: $M_{r} \sim -15.0$, $M_{i} \sim -14.9$ and $M_{z} \sim -14.8$ which correspond to $m_{g} \sim 21.08$, $m_{r} \sim 21.18$, $m_{i} \sim 21.28$ and $m_{z} \sim 21.38$ at $D_{L}=$172.2~Mpc (the SN and host distance). Thus, if a SMC like satellite was the host of SN~2024abvb, we would have detected it within the deep LIS imaging, effectively ruling out this possibility. 

\section{Discussion}
\label{sec:discussion}

\begin{figure}
	\includegraphics[width=\columnwidth]{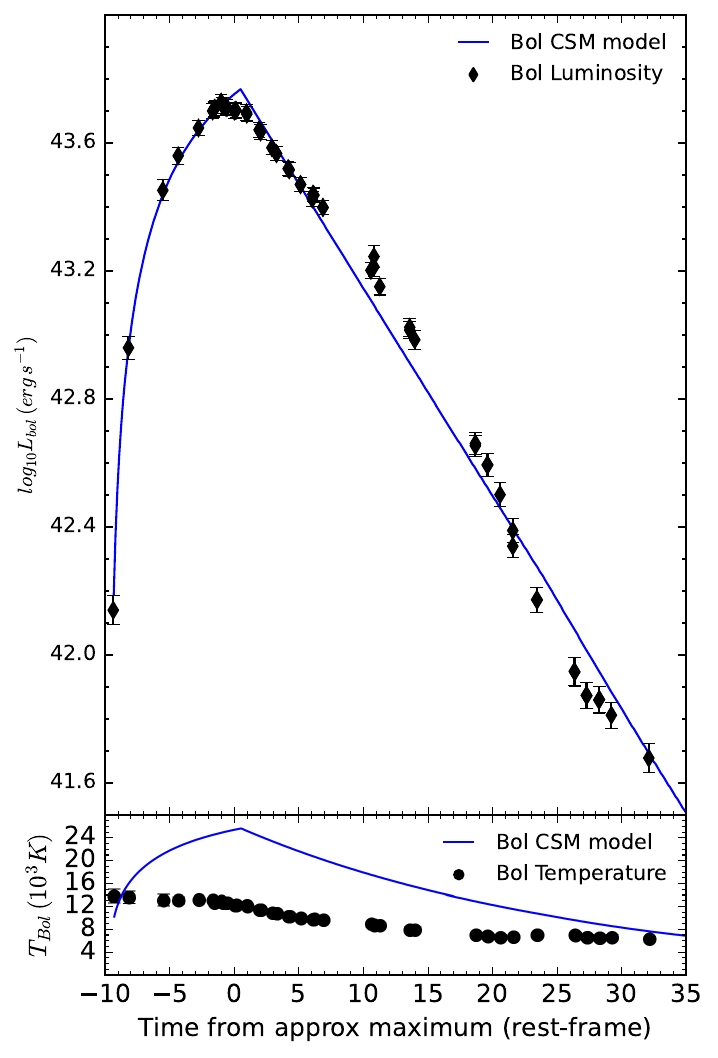}
    \caption{Bolometric light curves with CSM modelling. \textbf{Top Panel:} Bolometric light curves with a bolometric CSM model. We see a good fit ($\chi^{2} = 4.16$) of this model with our bolometric data.
    \textbf{Bottom Panel:} Black-body derived temperature and the CSM model.}
    \label{fig:bol_csm}
\end{figure}

\begin{figure}
	\includegraphics[width=\columnwidth]{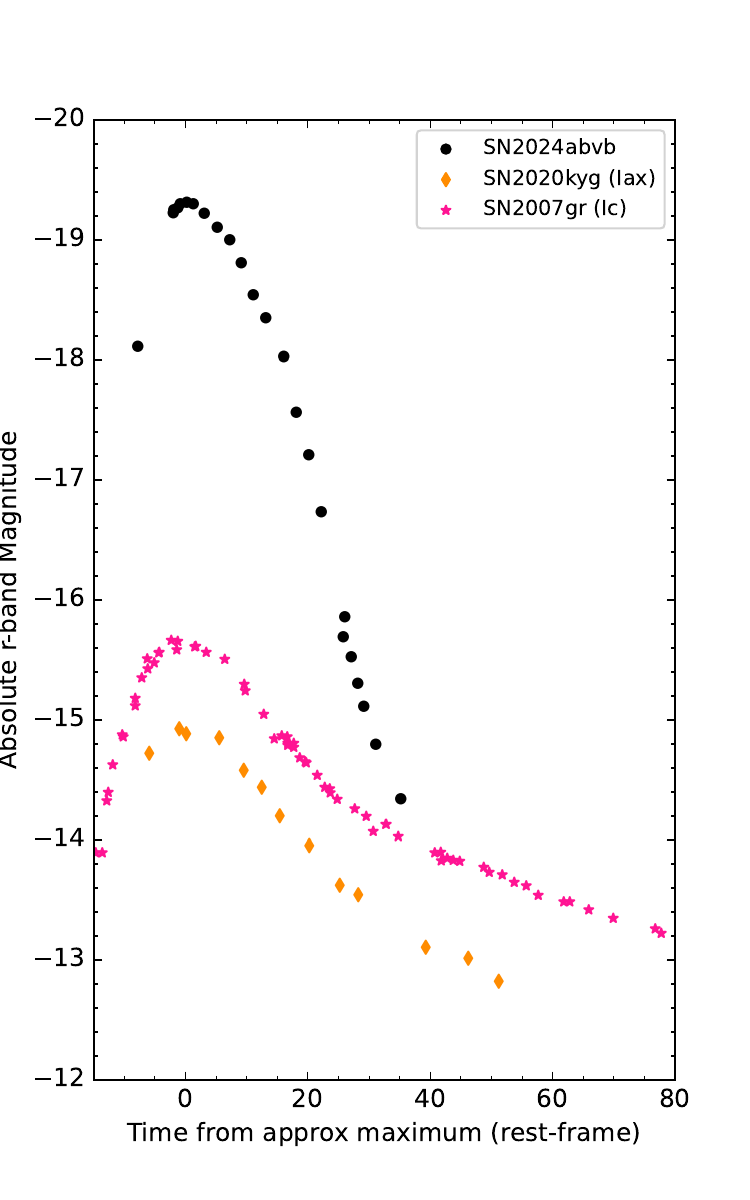}
    \caption{Comparison photometry in the $r$-band for SN~2024abvb, a Type~Iax SN (2020kyg, \citealp{2022MNRAS.511.2708S}) and a Type~Ic SN (2007gr, \citealp{2009A&A...508..371H}).}
    \label{fig:photometry_comp_sn}
\end{figure}

SN~2024abvb appears to be a peculiar event for several reasons: its offset from the host galaxy is atypical for its class (21.5~kpc); it shows C~{\sc ii}~$\lambda$5890 and He~{\sc i}~$\lambda$5876 in the spectra, with a velocity decrease from $\sim$1500~km~s$^{-1}$ to $\sim$800~km~s$^{-1}$; the unusually low ionization state of carbon or any other CNO element; a blackbody temperature up to a factor of two lower than that of other interacting Type~Ibn/Icn SNe during the first month of evolution.

The spectroscopic and photometric evolution of SN~2024abvb clearly indicates the presence of a dense CSM surrounding the progenitor. From the blackbody fit to the SED (see Section~\ref{sec:BLC}) and the bolometric luminosity, the peak radius is $2.1\times10^{15}$~cm, approximately $10^{3}$ times larger than that of a WR star \citep{2014A&A...565A..27H}, suggesting that the CSM is not gravitationally bound to the star. Following the formalism of \citet{2007ApJ...671L..17S}, the peak luminosity ($L_{\rm max}$) produced by the ejecta–CSM shock scales with the CSM shell mass, the pseudo-photospheric velocity ($v_{\rm ph}$), and the rise time ($t_{\rm max}$). Using $v_{\rm ph} = 1500$~km~s$^{-1}$ (Section~\ref{sec:vel}), $t_{\rm max} = 10.1$~d (Section~\ref{sec:LC}), and $L_{\rm max} = 2.9\times10^{43}$~erg~s$^{-1}$ (Section~\ref{sec:BLC}), we infer a CSM mass of roughly 1.7~$M_{\odot}$. Assuming a single-progenitor scenario, the aforementioned radius and a constant CSM velocity of 1500~km~s$^{-1}$, consistent with carbon-rich WR winds, the mass loss would have begun approximately 0.5~yr before the explosion. However, this would imply an enormous mass-loss rate ($\sim3.4$M$_{\odot}$~yr$^{-1}$), which has never been observed/derived for a WR. As a consequence, it is unlikely that the progenitor scenario involves a single WR star losing mass via wind.

To reproduce the bolometric light curve of SN~2024abvb under these assumptions, we employed a semi-analytical model based on \citet{2012ApJ...748...42C}. We investigated both shell and wind CSM configurations, finding that the shell model provides a better overall fit (lower $\chi^2$). We allowed for a contribution from $^{56}$Ni to the luminosity. The inner and outer power-law slopes of the SN ejecta density profile were fixed to $\delta = 2$ and $n = 9$, respectively. We adopted Thomson scattering as the dominant opacity source and assumed a helium-poor composition, giving $\kappa = 0.10$~cm$^{2}$~g$^{-1}$. We allowed for Ni$^{56}$ to contribute to the bolometric light curve fitting, but found that the best fit models required no Ni$^{56}$ contribution. Our best-fitting model, shown in the top panel of Figure~\ref{fig:bol_csm}, yields the following parameters: $M_{\mathrm{ej}} = 3.9~\mathrm{M}_{\odot}$, $M_{\mathrm{csm}} = 0.49~\mathrm{M}_{\odot}$ and $\mathrm{KE} = 0.3\times10^{51}~\mathrm{erg}$. The model reproduces the bolometric light curve relatively well ($\chi^{2} = 4.16$). The retrieved system properties point to a relatively low CSM mass, consistent with the photometric comparison in Figure~\ref{fig:photometry_comp}. The absence of Ni$^{56}$ suggests that the luminosity is dominated by ejecta–CSM interaction rather than radioactive heating. 
Additionally, we used the \textsc{MOSFiT}\footnote{MOSFiT is a Python package for fitting and estimating the parameters of transients via user-contributed transient models.} (The Modular Open Source Fitter for Transients) Python package to fit the bolometric light curve and its uncertainties. The model assumed a Type~Ic~SN with a CSM contribution and $\kappa=0.10$ cm$^2$ g$^{-1}$. The best fit (Figure \ref{fig:LC_ic} and \ref{fig:corner_plot}) returned a $M_{csm}=0.21~\mathrm{M_{\odot}}$ and a $M_{ej}=0.43~\mathrm{M_{\odot}}$ albeit with a much higher $\chi^2_{reduced}=175.2$ than the above semi-analytical model despite using the same formalism of \citet{2012ApJ...748...42C}.

Figure~\ref{fig:bol_csm} (bottom panel) shows a significant discrepancy between the temperature derived from blackbody fits and that predicted by the best-fitting CSM model. The temperature evolution for SN~2024abvb is also low when compared to SN~2019hgp (Figure~\ref{fig:bol}), indicating that SN~2024abvb exhibits intrinsically lower temperatures than other interacting Type~Ibn/Icn SNe.

The large projected distance of SN~2024abvb from its presumed host suggests two possible explanations. It could have been a long-lived star that was dynamically ejected \citep{Perets_2008,2018A&A...620A..48I,Jordan_2012}. Alternatively, it is possible for local high-SFR to be present within a low-SFR galaxy \citep{2009ApJ...707.1023V}, thus a short-lived progenitor that formed in situ is also possible. Because the spectra are dominated by CSM-interaction signatures, the underlying SN cannot be directly constrained spectroscopically.

For Completeness, before discussing the most compelling scenarios, we briefly address the possibility of a Type~Ia~SN being the underlying cause of SN~2024abvb. Type~Iax events are substantially less luminous than Type~Ia~SNe \citep{2014Natur.512...54M}, and are the only thermonuclear event with a lower luminosity than SN~2024abvb at all epochs.
In Type~Iax events, carbon typically appears in absorption \citep{Foley_2013} rather than emission, implying relatively cool material far from the progenitor system. Type~Iax SNe can exhibit carbon in their spectra, as their progenitors are C/O white dwarfs and some carbon may remain unburned during the explosion \citep{2006AJ....132..189J}. This is inconsistent with the dense, carbon-rich CSM located in close proximity to the progenitor inferred for SN~2024abvb.
Finally, merger events between two white dwarfs, particularly ONe+CO mergers as discussed by \citet{Wu_2024}, remain a plausible channel. Such mergers can produce bolometric light curves resembling those of Type Icn SNe such as SN~2019jc and SN~2021csp, and therefore could potentially explain SN~2024abvb. However, such a scenario should lead to an event somehow similar to SN~2005E \citep{2010Natur.465..322P} exhibiting calcium-rich spectral features. If SN~2024abvb arose from a similar channel we would likewise expect calcium lines to be present. The absence of calcium features in the spectra of SN~2024abvb does not obviously support this progenitor scenario.

\subsection{SN~2024abvb as a rare massive star in a region with a very low star formation rate}

Figure~\ref{fig:photometry_comp_sn} demonstrates that it is, in principle, possible to obscure a Type~Ic SN beneath the light curve of SN~2024abvb. 
The initial velocity of the C~{\sc ii}~$\lambda$5890 line, $\sim$1500~km~s$^{-1}$, is compatible with typical WR wind speeds. However, the subsequent decrease in velocity is difficult to explain in a scenario where the CSM is created by a constant wind, as unshocked wind material should not decelerate on the observed timescale \citep[e.g.][]{2019A&A...621A..92S, 2019A&A...625A..57H}. It is worth mentioning that a decrease in WR wind speed has been observed in WC9 stars, and explained via a line emission region transitioning from outer, lower density regions, to higher density regions, resulting in lower line velocities \citep{1971ApJ...164..275S}. However, such interpretation would not explain the weak He~{\sc i}, as WC9 stars typically show strong He~{\sc i} in their spectra, roughly equivalent to C~{\sc ii} in intensity \citep{1984ApJ...280..181T}.
This behaviour instead suggests that the CSM may have been intrinsically slower and that we never observed the unshocked component. Under this interpretation, a single-star progenitor pathway becomes unlikely given the inferred velocities. A binary scenario \citep[e.g.]{annurev:/content/journals/10.1146/annurev.astro.45.051806.110615} where one of the stars is a WR, by contrast, remains plausible as it naturally accommodates substantial mass stripping,
as well as significant mass loss shortly before explosion. This is broadly consistent with the order-of-magnitude estimates discussed above (see Section~\ref{sec:discussion}).

The isolated position of the progenitor could be explained through several channels. One possibility is that the star was a hypervelocity object  \citep[>800~km~s$^{-1}$,][]{Perets_2008} ejected from the host galaxy following an interaction with either the central supermassive black hole (SMBH) or an intermediate-mass black hole (IMBH) \citep{Perets_2008,2018A&A...620A..48I}. Alternatively, the star may have experienced an asymmetric eruptive event that imparted a substantial kick (a few hundred km~s$^{-1}$), displacing it from its natal environment \citep{Jordan_2012}. Although such eruptions are commonly associated with Type~Ia progenitors, an asymmetric outburst remains a viable mechanism for a massive star. Assuming that the massive star was ejected through the hypervelocity channel \citep[$\sim800$~km~s$^{-1}$]{Perets_2008}, and given a projected offset of 21.5~kpc, the travel time required to reach the explosion site is $\sim26.3$~Myr. This is much longer than the expected lifetimes of massive stars, which are $\sim5$~Myr \citep{2005A&A...429..581M, 2007ARA&A..45..177C}, rendering this scenario highly unlikely. The minimum projected separation between the SN position and the outermost spiral arm of the host galaxy is $\sim10.72$~kpc (0.231~arcmin), corresponding to a travel time of $\sim13.1$~Myr which is still more than twice the typical lifetime of stars in this mass range. Furthermore, \citet{2011MNRAS.414.3501E} argue that core-collapse SNe are unlikely to originate from progenitors located more than $\sim100$~pc from their birth sites.

Finally, it is possible, albeit rare, that the progenitor simply formed at its current location.

\subsection{SN~2024abvb as an ultra-stripped supernova}

Ultra-stripped SNe (USSNe, \citealp{Tauris_2013}) arise in binary systems composed of a compact object and a helium star. Such a configuration allows the system to be dynamically ejected from its host while still providing sufficient time for the helium star to evolve and explode as the observed SN. USSNe proceed through two main channels: iron core-collapse SNe (Fe CCSNe, \citealp{Tauris_2017}) and electron-capture SNe (EC SNe, \citealp{nomoto1987evolution}). While Fe CCSNe are essentially conventional core-collapse explosions of compact helium stars, EC SNe occur when the helium core collapses directly to a neutron star due to the loss of electron degeneracy pressure, triggered by electron captures in an ONeMg core. This collapse then produces a thermonuclear-like explosion \citep{Tauris_2017}.

The USSNe scenario naturally explains the presence of CSM. Binary stripping can deposit material in the immediate vicinity of the system with velocities lower than those expected for single WR-star winds. In such cases, stripping of the helium star exposes a nearly bare metal core, potentially accounting for the higher-order elements present in the spectra, which may trace the progenitor’s internal composition. USSNe are also predicted to synthesize little to no $^{56}$Ni, consistent with our bolometric light-curve modelling. In the EC SN channel, the ejecta mass is expected to be very low, which is compatible with the overall fast evolution of SN~2024abvb. However, the predicted ejecta masses for EC SNe ($\leq 0.2,M_{\odot}$; \citealp{tauris2015ultrastrippedsupernovaeprogenitorsfate}) are significantly smaller than the value inferred from our modelling.

\citet{moriya2025typeibnsupernovaeultrastripped} presented light-curve models for USSNe interacting with CSM, predicting a rapid rise to peak ($\sim 8$~days) followed by a relatively slow decline over $\sim 20$~days. These behaviours closely resemble the bolometric light curve of SN~2024abvb (Figure~\ref{fig:bol_csm}). Their models further suggest ejecta energies of $10^{50}$~erg, which align well with our best-fitting parameters and CSM masses of $\sim 0.2~M_{\odot}$, again comparable to our results. The primary caveat for adopting this interpretation is the ejecta mass, which in their models ($\sim$0.06~$M_{\odot}$) is substantially smaller than the value we infer (3.9~$M_{\odot}$) and only one order of magnitude smaller than the MCMC best fit, albeit retrieved with a high $\chi^2$ although we infer a much closer value from our MCMC driven light curves of $M_{ej}=0.43~\mathrm{M_{\odot}}$ while this is still larger, it is significantly closer in agreement than our previous estimate.
We note that \citet{hu2026sn2024abvbtypeicn} report $M_{ej}=0.12~\mathrm{M_\odot}$ from their \textsc{MOSFiT} fit. This discrepancy could arise from differences in the theoretical prescriptions implemented in \textsc{MOSFiT}, from different opacity assumptions, or from differences in the datasets as we have additional broadband imaging points during the rise and beyond 15 days after maximum light. Their \textsc{MOSFiT} posterior distribution shows a lower $\chi^2$ than ours, implying a better fit to the dataset. However, such codes assume spherical symmetry for both the ejecta and the CSM, whereas for SN~2024abvb \citet{theintelcollaboration2026nestedasymmetrichhecircumstellar} reported CSM asymmetries of up to 4\%. This indicates an additional source of uncertainty in outputs from simple semi-analytical models.

The remaining unresolved aspect across all progenitor scenarios considered is the unusually low observed temperature. This, however, can be explained through alternative processes, such as the formation of a cool/cold dense shell (CDS) \citep{2004MNRAS.352.1213C} with temperatures of $\sim10^4$K at the interface between the ejecta and the CSM. A denser CSM could lead to an optically thick environment in which radiation remains trapped, undergoing adiabatic losses that reduce the effective temperature despite strong instantaneous shock heating \citep{2004MNRAS.352.1213C,2006A&A...449..171N}. Aspherical CSM distributions can produce lower effective temperatures due to an increased radiating area \citep{2015MNRAS.449.4304D,Suzuki_2019}. Moreover, simulations of supernova explosions interacting with aspherical CSM distributions show that shock heating, density enhancement, and radiative diffusion depend strongly on the CSM geometry \citep{2020A&A...642A.214K}. \citet{theintelcollaboration2026nestedasymmetrichhecircumstellar} proposed that concentric toroidal shells could account for the relatively high asphericity ($P\sim4\%$) of SN~2024abvb, as well as for the structure observed in their higher-resolution spectra. Clumpy CSM structures can create similar outcomes, as the shock propagates through regions of varying density, generating temperature behaviour consistent with CDS-like cooling \citep{1994MNRAS.268..173C,Chandra_2018}. A decrease in the optical depth of the CSM \citep{theintelcollaboration2026nestedasymmetrichhecircumstellar}, driven by expansion, can expose cooler and deeper layers or modify the effective photosphere, leading to a lower inferred temperature \citep{2021A&G....62.5.34N}. Such effects are more likely in binary scenarios, where asphericity is expected to be more pronounced, and can therefore result in systematically lower observed temperatures compared to symmetric configurations.

\section{Summary and Conclusions}
\label{sec:conclusion}

We have presented spectroscopic and photometric observations of SN~2024abvb, a Type~Ibn/Icn event that exhibits several properties distinct from the broader Type~Ibn/Icn SN population, including an unusually large projected separation from its presumed host galaxy (21.5~kpc). The combined spectroscopic and photometric dataset highlights how atypical SN~2024abvb is relative to other hydrogen- and helium-deficient interacting supernovae.

SN~2024abvb displays a $g$-band rise time notably longer than that of typical Type~Icn, and Ibn, events. The ATLAS $o$-band rise time is also marginally extended relative to other Type~Icns. All photometric bands exhibit linear post-peak declines, with the steepest fading observed in the UV and blue optical filters, and the shallowest in the NIR. This behaviour is consistent with sustained high-energy photon reprocessing.

In the $r$-band, SN~2024abvb declines at a rate comparable to other Type~Icn events but peaks at a higher luminosity. This suggests a similar low-mass CSM environment and progenitor pathway as for SN~2019hgp, albeit with a slightly higher peak magnitude ($M_r \approx -19.2$~mag), indicative of a more efficient conversion of shock energy into radiation.

The bolometric light curve of SN~2024abvb declines approximately linearly after peak, but the peak black-body temperature is substantially lower than that found in other interacting Type~Ibn/Icn SNe. The fractional flux evolution confirms that UV emission dominates at early times, with optical flux rising to match the UV contribution as the ejecta–CSM interaction proceeds. The NIR fraction initially decreases and then stabilises slightly above its early value, suggesting minimal dust formation or dust formation occurring after the epochs probed by our last NIR observations. The overall bolometric evolution resembles the CSM-interaction USSN models of \citet{moriya2025typeibnsupernovaeultrastripped}, with the exception of the larger ejecta mass inferred for SN~2024abvb.

Spectroscopically, SN~2024abvb exhibits an initially blue continuum with multiple narrow P-Cygni features, dominated by C~{\sc ii}~$\lambda5890$ in emission. The continuum temperature decreases with time, reaching a plateau after $\sim$20~days, likely due to metal-line blanketing. Weak He~{\sc i}~$\lambda5876$ absorption is detected in the earliest medium-resolution spectra, and the same spectra reveal a broad red-wing shoulder associated with C~{\sc ii}~$\lambda5890$, further suggesting heavy metal-line blanketing or possibly an asymmetric CSM configuration.

O~{\sc ii}~$\lambda4651$ is present in the early-time spectra, broadens at intermediate epochs, and disappears at later times. In contrast, C~{\sc ii}~$\lambda5890$ persists throughout the observed period, transitioning to a P-Cygni profile in both low- and medium-resolution spectra after $\sim$20~days. The medium-resolution data also show H~{\sc i}~$\lambda4861$, O~{\sc ii}~$\lambda4651$, and Fe~{\sc ii}~$\lambda5363$ with discernible P-Cygni features even at late times.

Comparisons with other Type~Icn and Type~Ibn SNe reinforce the uniqueness of SN~2024abvb. While some similarities exist, the spectra and photometric evolution strongly disfavour a Type~Ibn classification and instead support a carbon-rich, low-mass CSM environment consistent with an Type~Ibn/Icn SN.

The expansion velocities of SN~2024abvb peak at values compatible with canonical WR wind speeds. However, the subsequent decline is inconsistent with expectations for an unshocked WR-like wind or shell. This suggests that the CSM may have been produced either by a discrete eruptive episode or by binary interaction. In an eruptive scenario, the implied mass-loss rate would be high enough to dominate the local density structure, while a steady WR wind alone would struggle to reproduce the observed CSM properties. The velocities are comparable to those of other interacting SNe but point toward a progenitor pathway more complex than that of a single WR star.

It is unlikely that SN~2024abvb originated from a thermonuclear (Type~Ia) explosion; the corresponding light curves are incompatible with those observed. A scenario involving a rare massive star formed in a low-star-formation region also appears improbable given the extremely low local SFR. The most plausible explanation is that SN~2024abvb was an USSNe, a scenario capable of accounting for most of its observed properties and naturally explaining the extreme separation from the host. However, this interpretation does not reconcile the discrepancy between the ejecta mass inferred for SN~2024abvb and the significantly lower ejecta masses predicted by current USSNe models (e.g.\ \citealp{Tauris_2017, moriya2025typeibnsupernovaeultrastripped}).

Ultimately, advancing our understanding of interacting SNe will require a larger sample of well-observed events, with dense photometric coverage and, crucially, higher-resolution spectroscopy and spectropolarimetry to probe potential asymmetries in their ejecta and circumstellar environments. Such datasets are essential for disentangling the diverse progenitor channels that give rise to interacting explosions and for clarifying the roles of single and binary evolution in shaping their observable properties. With improved observational constraints, we can move toward a more complete and physically grounded picture of the late-stage evolution of massive stars and the mechanisms that produce these rare and intriguing transients.
\section*{Acknowledgements}
CI gratefully acknowledges the support received from the MERAC Foundation. T.E.M.B. is funded by Horizon Europe ERC grant no. 101125877. 
J.R.F. is supported by the U.S. National Science Foundation (NSF) Graduate Research Fellowship Program under grant 2139319.
CPG acknowledges financial support from the Secretary of Universities and Research (Government of Catalonia) and by the Horizon 2020 Research and Innovation Programme of the European Union under the Marie Sk\l{}odowska-Curie and the Beatriu de Pin\'os 2021 BP 00168 programme, from the Spanish Ministerio de Ciencia e Innovaci\'on (MCIN) and the Agencia Estatal de Investigaci\'on (AEI) 10.13039/501100011033 under the PID2023-151307NB-I00 SNNEXT project, from Centro Superior de Investigaciones Cient\'ificas (CSIC) under the PIE project 20215AT016 and the program Unidad de Excelencia Mar\'ia de Maeztu CEX2020-001058-M, and from the Departament de Recerca i Universitats de la Generalitat de Catalunya through the 2021-SGR-01270 grant. 
TLK acknowledges support via a Warwick Astrophysics prize post-doctoral fellowship made possible thanks to a generous philanthropic donation. 
GL was supported by a research grant (VIL60862) from VILLUM FONDEN. 
SM is funded by Leverhulme Trust grant RPG-2023-240. 
MP acknowledges support from a UK Research and Innovation Fellowship (MR/T020784/1 and UKRI1062). 
Based on observations collected at Copernico and Schmidt telescopes (Asiago Mount Ekar, Italy) of the INAF -- Osservatorio Astronomico di Padova.
AR acknowledges financial support from the GRAWITA Large Program Grant (PI P. D’Avanzo). 
AR, AP, GV acknowledge financial support from the PRIN-INAF 2022 ``Shedding light on the nature of gap transients: from the observations to the models''
BW acknowledges the UKRI’s STFC studentship grant funding, project reference ST/X508871/1
FEB acknowledges support from ANID-Chile BASAL CATA FB210003, FONDECYT Regular 1241005, and Millennium Science Initiative, AIM23-0001.
T.-W.C. acknowledges support from the Ministry of Education Yushan Fellow Program (MOE-111-YSFMS-0008-001-P1) and from the National Science and Technology Council, Taiwan (NSTC 114-2112-M-008-021-MY3).
G.V. and A.R. acknowledge financial support from the SOXS project (PI S. Campana).
This article is also based on observations obtained from the La Silla Observatory with the REM telescope, under the program REM AOT47-37 (ID 49337, PI: G. Valerin).
KAB is supported by an LSST-DA Catalyst Fellowship; this publication was thus made possible through the support of Grant 62192 from the John Templeton Foundation to LSST-DA. JDL acknowledges support from a UK Research and Innovation Future Leaders Fellowship (grant references MR/T020784/1 and UKRI1062).

Based on observations collected at the European Organisation for Astronomical Research in the Southern Hemisphere, Chile, as part of ePESSTO+ (the advanced Public ESO Spectroscopic Survey for Transient Objects Survey – PI: Inserra). ePESSTO+ observations were obtained under ESO program ID 112.25JQ. This work makes use of data from the Asteroid Terrestrial-impact Last Alert System (ATLAS) project. The ATLAS project is primarily funded by NASA grants.
The Gravitational-wave Optical Transient Observer (GOTO) project acknowledges the support of the Monash-Warwick Alliance; University
of Warwick; Monash University; University of Sheffield; University of Leicester; Armagh Observatory \& Planetarium; the National
Astronomical Research Institute of Thailand (NARIT); Instituto de
Astrofísica de Canarias (IAC); University of Portsmouth; University of Turku; University of Manchester and the UK Science and
Technology Facilities Council (STFC, grant numbers ST/T007184/1,
ST/T003103/1 and ST/T000406/1).
This article includes observations made in the Two-meter Twin Telescope (TTT) in the Teide Observatory of the IAC, that Light Bridges operates in the Island of Tenerife, Canary Islands (Spain). The Observing Time Rights (DTO) used for this research were provided by Light Bridges, SL.
This paper uses data that were obtained by The Legacy Surveys: the Dark Energy Camera Legacy Survey (DECaLS; NOAO Proposal ID \# 2014B-0404; PIs: David Schlegel and Arjun Dey), the Beijing-Arizona Sky Survey (BASS; NOAO Proposal ID \# 2015A-0801; PIs: Zhou Xu and Xiaohui Fan), and the Mayall z-band Legacy Survey (MzLS; NOAO Proposal ID \# 2016A-0453; PI: Arjun Dey). DECaLS, BASS and MzLS together include data obtained, respectively, at the Blanco telescope, Cerro Tololo Inter-American Observatory, National Optical Astronomy Observatory (NOAO); the Bok telescope, Steward Observatory, University of Arizona; and the Mayall telescope, Kitt Peak National Observatory, NOAO. NOAO is operated by the Association of Universities for Research in Astronomy (AURA) under a cooperative agreement with the National Science Foundation. Please see http://legacysurvey.org for details regarding the Legacy Surveys. BASS is a key project of the Telescope Access Program (TAP), which has been funded by the National Astronomical Observatories of China, the Chinese Academy of Sciences (the Strategic Priority Research Program "The Emergence of Cosmological Structures" Grant No. XDB09000000), and the Special Fund for Astronomy from the Ministry of Finance. The BASS is also supported by the External Cooperation Program of Chinese Academy of Sciences (Grant No. 114A11KYSB20160057) and Chinese National Natural Science Foundation (Grant No. 11433005). The Legacy Surveys imaging of the DESI footprint is supported by the Director, Office of Science, Office of High Energy Physics of the U.S. Department of Energy under Contract No. DE-AC02-05CH1123, and by the National Energy Research Scientific Computing Center, a DOE Office of Science User Facility under the same contract; and by the U.S. National Science Foundation, Division of Astronomical Sciences under Contract No.AST-0950945 to the National Optical Astronomy Observatory.
The VISTA Hemisphere Survey data products served at Astro Data Lab are based on observations collected at the European Organisation for Astronomical Research in the Southern Hemisphere under ESO programme 179.A-2010, and/or data products created thereof.
Pan-STARRS is a project of the Institute for Astronomy of the University of Hawaii, and is supported by the NASA SSO Near Earth Observation Program under grants 80NSSC18K0971, NNX14AM74G, NNX12AR65G, NNX13AQ47G, NNX08AR22G, 80NSSC21K1572 and by the State of Hawaii. 
This work makes use of observations from the Las Cumbres Observatory network. The LCO team is supported by NSF grants AST-2308113 and AST-1911151. 
\section*{Data Availability}

Photometry data are listed in the appendix. All spectra are available on WISeREP (https://www.wiserep.org).



\bibliographystyle{mnras}
\bibliography{bibliography} 




\appendix

\section{Data tables}

\begin{figure}
    \includegraphics[width=\columnwidth]{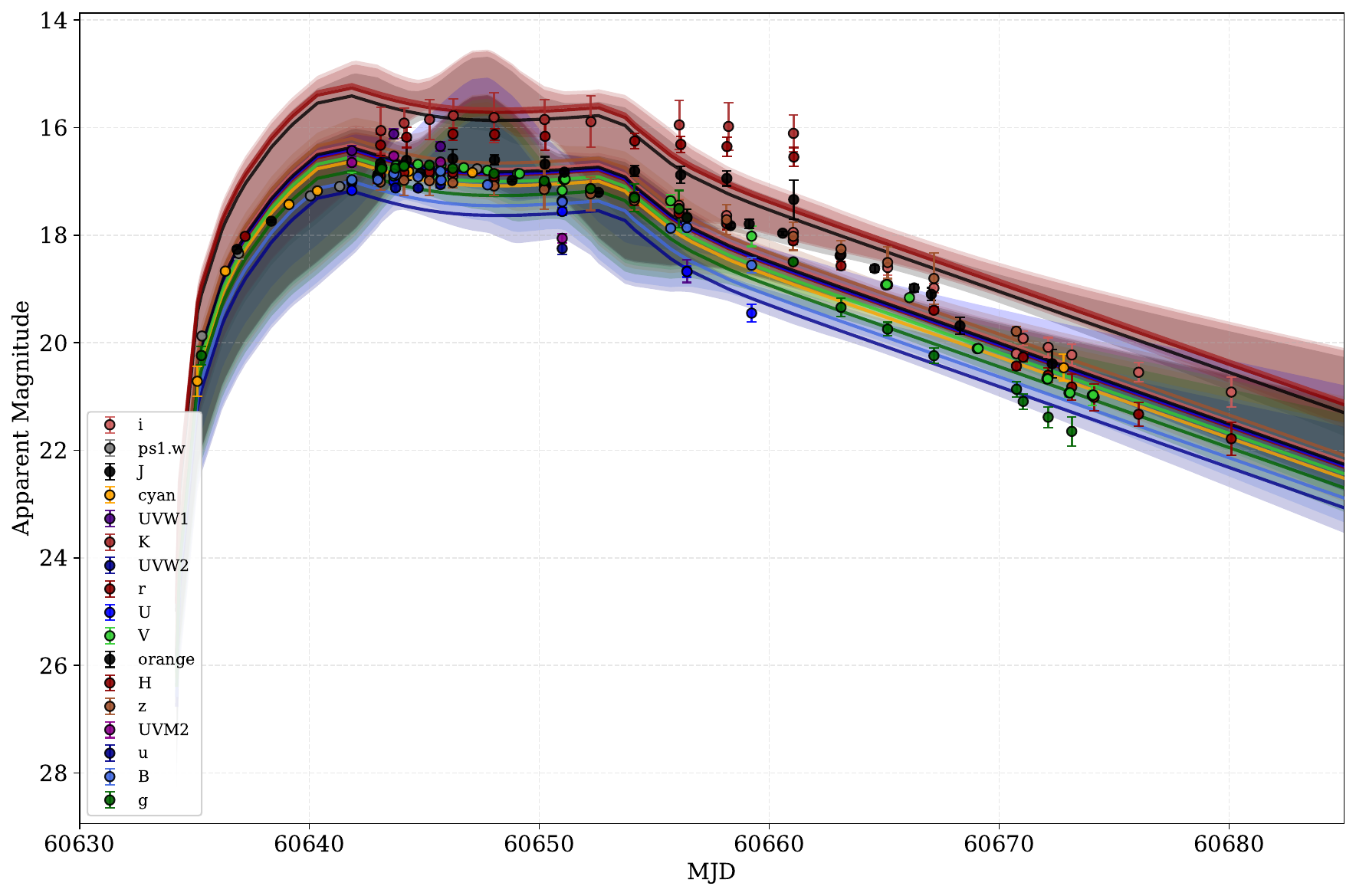}
    \caption{}
    \label{fig:LC_ic}
\end{figure}

\begin{figure}
    \includegraphics[width=\columnwidth]{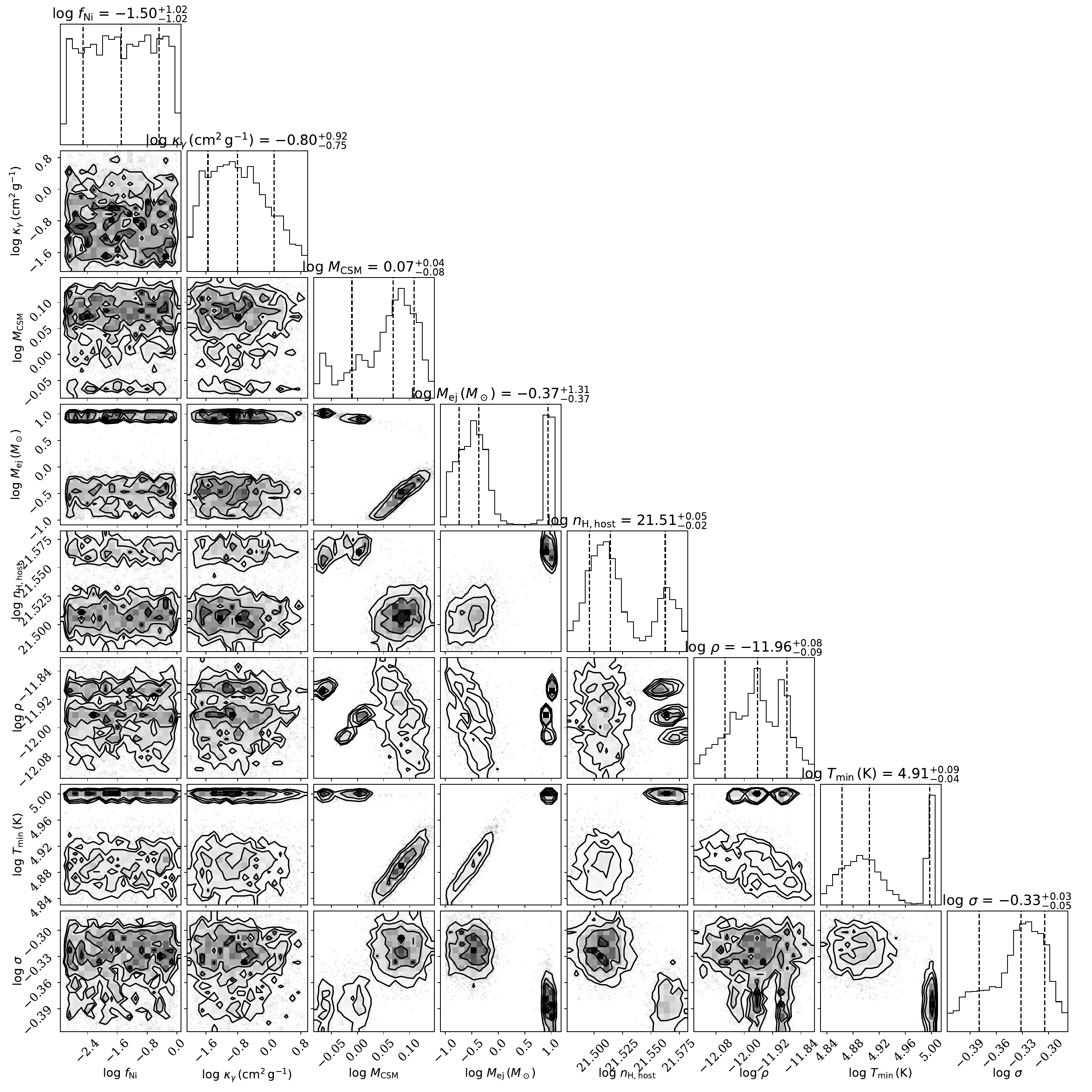}
    \caption{}
    \label{fig:corner_plot}
\end{figure}

\begin{table*}
\caption{Table of all spectral observations.}
\label{tab:spectra_obs}
\begin{tabular}{lllllrl}
\hline
Date of obs             & MJD      & Rest frame phase & Instrument & Grism          & Slit & Resolution      \\
 & & (w.r.t maximum in days) & & & & (\AA)\\ \hline
2024-11-27 & 60641.90 & -3.2 & NOT/ALFOSC & Gr4 & 1.0" & 16.2     \\
2024-11-29 & 60643.08 & -2.1 & NTT/EFOSC2 & Gr11,Gr16,Gr18 & 1.0" & 17.16, 17.29, 8.19     \\
2024-11-30 & 60644.06 & -1.1 & NTT/EFOSC2 & Gr11,Gr16,Gr18 & 1.0" & 17.16, 17.29, 8.19     \\
2024-12-02 & 60646.22 & 1.1 & FLOYDS &  & 2.0" &   17.0   \\
2024-12-05 & 60649.05 & 3.9 & NTT/EFOSC2 & Gr11,Gr16,Gr18 & 1.0" & 17.16, 17.29, 8.19     \\
2024-12-07 & 60651.05 & 5.9 & NTT/EFOSC2 & Gr11,Gr16,Gr18 & 1.0" & 17.16, 17.29, 8.19     \\
2024-12-08 & 60652.21 & 7.1 & FLOYDS &  & 2.0" & 17.0 \\
2024-12-19 & 60663.27 & 18.1 & FLOYDS &  & 2.0" & 17.0 \\
2024-12-21 & 60665.06 & 19.9 & NTT/EFOSC2 & Gr11,Gr16      & 1.0" & 17.16, 17.29     \\
2024-12-22 & 60666.10  & 21.0 & NTT/EFOSC2 & Gr18           & 1.0" & 8.19     \\
2024-12-25 & 60669.04 & 23.9 & NTT/EFOSC2 & Gr11,Gr16      & 1.0" & 17.16, 17.29    \\
2024-12-29 & 60673.03 & 27.9 & NTT/EFOSC2 & Gr16           & 1.0" & 17.29 \\
2024-12-30 & 60674.04 & 28.9 & NTT/EFOSC2 & Gr18           & 1.0" & 8.19 \\ \hline
\end{tabular}
\end{table*}

\clearpage
\onecolumn
\begin{longtable}{cclll}
\caption{A table displaying all photometric data used.}
\label{tab:full_phot}\\
\hline
MJD      & Band   & mag & err   & marker             \\ \hline
\endfirsthead
\multicolumn{5}{c}%
{{\bfseries Table \thetable\ continued from previous page}} \\
\hline
MJD      & Band   & mag & err   & marker             \\ \hline
\endhead
\hline
\endfoot
\endlastfoot
60641.84 & B      & 16.970   & 0.080 & UVOT               \\
60642.97 & B      & 16.971   & 0.100 & NOT                \\
60643.67 & B      & 16.860   & 0.060 & UVOT               \\
60643.77 & B      & 16.876   & 0.027 & 67/91,Moravian     \\
60644.72 & B      & 16.916   & 0.025 & 67/91,Moravian     \\
60644.84 & B      & 16.835   & 0.013 & LCO                \\
60644.84 & B      & 16.827   & 0.014 & LCO                \\
60645.70 & B      & 16.810   & 0.060 & UVOT               \\
60645.73 & B      & 16.977   & 0.032 & 67/91,Moravian     \\
60646.72 & B      & 16.022   & 0.036 & 67/91,Moravian     \\
60646.83 & B      & 16.937   & 0.015 & LCO                \\
60646.83 & B      & 16.949   & 0.014 & LCO                \\
60647.75 & B      & 17.065   & 0.058 & 67/91,Moravian     \\
60648.10 & B      & 17.054   & 0.024 & LCO                \\
60648.11 & B      & 17.043   & 0.024 & LCO                \\
60650.04 & B      & 17.258   & 0.017 & LCO                \\
60650.04 & B      & 17.266   & 0.018 & LCO                \\
60650.99 & B      & 17.380   & 0.080 & UVOT               \\
60651.85 & B      & 17.348   & 0.017 & LCO                \\
60651.85 & B      & 17.351   & 0.017 & LCO                \\
60655.71 & B      & 17.871   & 0.068 & 67/91,Moravian     \\
60655.94 & B      & 17.811   & 0.032 & LCO                \\
60655.94 & B      & 17.855   & 0.029 & LCO                \\
60656.42 & B      & 17.860   & 0.090 & UVOT               \\
60658.82 & B      & 18.426   & 0.035 & LCO                \\
60658.83 & B      & 18.442   & 0.036 & LCO                \\
60659.23 & B      & 18.560   & 0.140 & UVOT               \\
60664.10 & B      & 19.770   & 0.029 & LCO                \\
60664.11 & B      & 19.771   & 0.031 & LCO                \\
60667.11 & B      & 20.662   & 0.120 & LCO                \\
60667.11 & B      & 20.466   & 0.092 & LCO                \\
60669.71 & B      & >20.120   & -    & 67/91,Moravian     \\
60675.04 & B      & 21.518   & 0.154 & LCO                \\
60675.04 & B      & 22.408   & 0.282 & LCO                \\
60678.05 & B      & 22.076   & 0.248 & LCO                \\
60690.05 & B      & 20.967   & 0.222 & LCO                \\
60693.04 & B      & 22.054   & 0.429 & LCO                \\
60693.04 & B      & 22.311   & 0.416 & LCO                \\
60683.81 & B      & >21.711	 & -	 & Liverpool          \\

60641.84	&	 V      	&	16.98	&	0.15	&	 UVOT               \\
60642.97	&	 V      	&	16.906	&	0.1	&	 NOT                \\
60643.08	&	 V      	&	16.75	&	0.054	&	 NTT                \\
60643.08	&	 V      	&	16.746	&	0.023	&	 NTT                \\
60643.15	&	 V      	&	16.741	&	0.042	&	 NTT                \\
60643.67	&	 V      	&	16.9	&	0.12	&	 UVOT               \\
60643.77	&	 V      	&	16.705	&	0.027	&	 67/91,Moravian     \\
60644.05	&	 V      	&	16.712	&	0.042	&	 NTT                \\
60644.06	&	 V      	&	16.714	&	0.036	&	 NTT                \\
60644.14	&	 V      	&	16.705	&	0.04	&	 NTT                \\
60644.72	&	 V      	&	16.685	&	0.024	&	 67/91,Moravian     \\
60644.84	&	 V      	&	16.718	&	0.016	&	 LCO                \\
60644.85	&	 V      	&	16.723	&	0.018	&	 LCO                \\
60645.7	&	 V      	&	16.84	&	0.1	&	 UVOT               \\
60645.72	&	 V      	&	16.704	&	0.023	&	 67/91,Moravian     \\
60646.71	&	 V      	&	16.745	&	0.035	&	 67/91,Moravian     \\
60646.83	&	 V      	&	16.794	&	0.018	&	 LCO                \\
60646.83	&	 V      	&	16.797	&	0.018	&	 LCO                \\
60647.74	&	 V      	&	16.796	&	0.037	&	 67/91,Moravian     \\
60648.11	&	 V      	&	16.88	&	0.065	&	 LCO                \\
60648.11	&	 V      	&	16.879	&	0.046	&	 LCO                \\
60649.05	&	 V      	&	16.856	&	0.024	&	 NTT                \\
60649.14	&	 V      	&	16.857	&	0.044	&	 NTT                \\
60650.04	&	 V      	&	17.008	&	0.015	&	 LCO                \\
60650.05	&	 V      	&	16.982	&	0.015	&	 LCO                \\
60650.99	&	 V      	&	17.17	&	0.13	&	 UVOT               \\
60651.05	&	 V      	&	16.957	&	0.02	&	 NTT                \\
60651.08	&	 V      	&	16.971	&	0.026	&	 NTT                \\
60651.12	&	 V      	&	16.96	&	0.038	&	 NTT                \\
60651.85	&	 V      	&	17.093	&	0.019	&	 LCO                \\
60651.85	&	 V      	&	17.108	&	0.021	&	 LCO                \\
60655.7	&	 V      	&	17.361	&	0.109	&	 67/91,Moravian     \\
60655.95	&	 V      	&	17.121	&	0.206	&	 LCO                \\
60655.95	&	 V      	&	16.787	&	0.201	&	 LCO                \\
60656.42	&	 V      	&	17.67	&	0.15	&	 UVOT               \\
60658.83	&	 V      	&	17.837	&	0.027	&	 LCO                \\
60658.83	&	 V      	&	17.925	&	0.029	&	 LCO                \\
60659.23	&	 V      	&	18.02	&	0.19	&	 UVOT               \\
60664.11	&	 V      	&	18.821	&	0.028	&	 LCO                \\
60664.11	&	 V      	&	18.789	&	0.028	&	 LCO                \\
60665.06	&	 V      	&	18.926	&	0.034	&	 NTT                \\
60665.11	&	 V      	&	18.919	&	0.025	&	 NTT                \\
60666.09	&	 V      	&	19.162	&	0.04	&	 NTT                \\
60667.12	&	 V      	&	19.508	&	0.064	&	 LCO                \\
60667.12	&	 V      	&	19.77	&	0.07	&	 LCO                \\
60669.04	&	 V      	&	20.111	&	0.064	&	 NTT                \\
60669.09	&	 V      	&	20.108	&	0.078	&	 NTT                \\
60669.7	&	 V      	&	 >19.947   	&	 -    	&	 67/91,Moravian     \\
60672.07	&	 V      	&	20.672	&	0.071	&	 NTT                \\
60672.11	&	 V      	&	20.673	&	0.066	&	 NTT                \\
60673.03	&	 V      	&	20.932	&	0.082	&	 NTT                \\
60673.08	&	 V      	&	20.933	&	0.076	&	 NTT                \\
60674.04	&	 V      	&	20.981	&	0.18	&	 NTT                \\
60674.09	&	 V      	&	20.972	&	0.094	&	 NTT                \\
60675.05	&	 V      	&	21.159	&	0.152	&	 LCO                \\
60675.05	&	 V      	&	21.422	&	0.185	&	 LCO                \\
60678.06	&	 V      	&	21.496	&	0.321	&	 LCO                \\
60678.06	&	 V      	&	21.391	&	0.281	&	 LCO                \\
60683.82	&	 V      	&	 >20.399   	&	 -    	&	 Liverpool          \\
60687.06	&	 V      	&	21.037	&	0.321	&	 LCO                \\
60690.06	&	 V      	&	21.049	&	0.38	&	 LCO                \\

60641.84	&	 U      	&	17.17	&	0.07	&	 UVOT               \\
60643.67	&	 U      	&	17	&	0.06	&	 UVOT               \\
60643.75	&	 U      	&	17.121	&	0.05	&	 67/91,Moravian     \\
60644.73	&	 U      	&	17.124	&	0.05	&	 67/91,Moravian     \\
60645.7	&	 U      	&	16.98	&	0.05	&	 UVOT               \\
60650.99	&	 U      	&	17.56	&	0.07	&	 UVOT               \\
60656.42	&	 U      	&	18.68	&	0.1	&	 UVOT               \\
60659.23	&	 U      	&	19.45	&	0.17	&	 UVOT               \\

60614.36	&	 g      	&	 >21.124   	&	 -    	&	 ZTF                \\
60622.19	&	 g      	&	 >20.420   	&	 -    	&	 ZTF                \\
60632.13	&	 g      	&	 >20.180   	&	 -    	&	 ZTF                \\
60634.25	&	 g      	&	 >20.764   	&	 -    	&	 ZTF                \\
60635.3	&	 g      	&	20.238	&	0.174	&	 ZTF                \\
60637.23	&	 g      	&	 >20.239   	&	 -    	&	 ZTF                \\
60639.31	&	 g      	&	 >16.841   	&	 -    	&	 ZTF                \\
60642.82    &    g          &   16.910  &   0.016   &    TTT \\
60642.98	&	 g      	&	16.699	&	0.013	&	 LCO                \\
60642.99    &    g          &   16.892  &   0.016   &    TTT \\
60643.09	&	 g      	&	16.781	&	0.182	&	 REM                \\
60643.15	&	 g      	&	16.767	&	0.041	&	 ZTF                \\
60643.73	&	 g      	&	16.757	&	0.022	&	 67/91,Moravian-DIF \\
60643.97    &    g          &   16.830  &   0.022   &    TTT \\
60644.11	&	 g      	&	16.718	&	0.027	&	 REM                \\
60644.81	&	 g      	&	16.665	&	0.011	&	 LCO                \\
60644.82    &    g          &   16.870  &   0.016   &    TTT \\
60645.01    &    g          &   16.855  &   0.019   &    TTT \\
60645.22	&	 g      	&	16.7	&	0.041	&	 REM                \\
60646.23	&	 g      	&	16.76	&	0.056	&	 REM                \\
60646.79	&	 g      	&	16.787	&	0.012	&	 LCO                \\
60646.82    &    g          &   16.946  &   0.026   &    TTT \\
60647.82    &    g          &   17.036  &   0.016   &    TTT \\
60647.99    &    g          &   17.024  &   0.018   &    TTT \\
60648.03	&	 g      	&	16.855	&	0.057	&	 REM                \\
60649.08	&	 g      	&	16.94	&	0.013	&	 LCO                \\
60650.21	&	 g      	&	16.992	&	0.05	&	 REM                \\
60650.85	&	 g      	&	17.124	&	0.014	&	 LCO                \\
60652.23	&	 g      	&	17.137	&	0.065	&	 REM                \\
60654.13	&	 g      	&	17.306	&	0.253	&	 REM                \\
60656.07	&	 g      	&	17.514	&	0.344	&	 REM                \\
60656.46	&	 g      	&	17.683	&	0.032	&	 LCO                \\
60658.46	&	 g      	&	18.108	&	0.039	&	 LCO                \\
60660.46	&	 g      	&	18.526	&	0.058	&	 LCO                \\
60661.04	&	 g      	&	18.497	&	0.05	&	 REM                \\
60662.92	&	 g      	&	18.991	&	0.063	&	 LCO                \\
60663.12	&	 g      	&	19.343	&	0.172	&	 REM                \\
60664.46	&	 g      	&	19.42	&	0.056	&	 LCO                \\
60665.14	&	 g      	&	19.748	&	0.127	&	 REM                \\
60666.81	&	 g      	&	20.193	&	0.115	&	 LCO                \\
60667.16	&	 g      	&	20.242	&	0.144	&	 REM                \\
60668.47	&	 g      	&	20.51	&	0.15	&	 LCO                \\
60670.75	&	 g      	&	20.866	&	0.141	&	 AFOSC              \\
60671.03	&	 g      	&	21.092	&	0.14	&	 REM                \\
60672.13	&	 g      	&	21.385	&	0.194	&	 REM                \\
60672.46	&	 g      	&	21.597	&	0.324	&	 LCO                \\
60673.15	&	 g      	&	21.649	&	0.267	&	REM                 \\
60675.06	&	 g      	&	21.308	&	0.272	&	 LCO                \\
60676.06	&	 g      	&	 >21.623   	&	-	&	 REM                \\
60680.09	&	 g      	&	 >21.646   	&	-	&	 REM                \\
60683.82	&	 g      	&	 >20.642   	&	-	&	 Liverpool          \\

57406.24	&	 r      	&	 >22.773   	&	 -    	&	 PS1                \\
60614.2	&	 r      	&	 >20.779   	&	 -    	&	 ZTF                \\
60616.33	&	 r      	&	 >21.216   	&	 -    	&	 ZTF                \\
60616.33	&	 r      	&	21.72	&	0.462	&	 ZTF                \\
60619.17	&	 r      	&	 >20.953   	&	 -    	&	 ZTF                \\
60620.14	&	 r      	&	 >20.294   	&	 -    	&	 ZTF                \\
60622.17	&	 r      	&	 >20.880   	&	 -    	&	 ZTF                \\
60625.33	&	 r      	&	 >19.996   	&	 -    	&	 ZTF                \\
60632.23	&	 r      	&	 >20.951   	&	 -    	&	 ZTF                \\
60634.23	&	 r      	&	 >20.410   	&	 -    	&	 ZTF                \\
60635.17	&	 r      	&	 >20.817   	&	 -    	&	 ZTF                \\
60637.2	&	 r      	&	18.023	&	0.021	&	 ZTF                \\
60642.97	&	 r      	&	16.91	&	0.121	&	 NOT                \\
60642.98	&	 r      	&	16.851	&	0.017	&	 LCO                \\
60643.09	&	 r      	&	16.885	&	0.029	&	 REM                \\
60643.74	&	 r      	&	16.867	&	0.024	&	 67/91,Moravian-DIF \\
60644.11	&	 r      	&	16.836	&	0.063	&	 REM                \\
60644.81	&	 r      	&	16.822	&	0.012	&	 LCO                \\
60645.22	&	 r      	&	16.824	&	0.067	&	 REM                \\
60646.23	&	 r      	&	16.837	&	0.034	&	 REM                \\
60646.79	&	 r      	&	16.846	&	0.011	&	 LCO                \\
60648.03	&	 r      	&	16.914	&	0.022	&	 REM                \\
60649.08	&	 r      	&	17.024	&	0.015	&	 LCO                \\
60650.21	&	 r      	&	17.03	&	0.029	&	 REM                \\
60650.85	&	 r      	&	17.051	&	0.014	&	 LCO                \\
60652.23	&	 r      	&	17.137	&	0.036	&	 REM                \\
60654.13	&	 r      	&	17.329	&	0.121	&	 REM                \\
60656.07	&	 r      	&	17.594	&	0.058	&	 REM                \\
60656.46	&	 r      	&	17.514	&	0.03	&	 LCO                \\
60658.14	&	 r      	&	17.785	&	0.112	&	 REM                \\
60658.46	&	 r      	&	17.682	&	0.028	&	 LCO                \\
60660.46	&	 r      	&	18.109	&	0.039	&	 LCO                \\
60661.04	&	 r      	&	18.106	&	0.062	&	 REM                \\
60662.92	&	 r      	&	18.461	&	0.042	&	 LCO                \\
60663.12	&	 r      	&	18.57	&	0.08	&	 REM                \\ 
60664.46	&	 r      	&	18.818	&	0.042	&	 LCO                \\
60665.14	&	 r      	&	18.922	&	0.067	&	 REM                \\
60666.81	&	 r      	&	19.232	&	0.063	&	 LCO                \\
60667.16	&	 r      	&	19.397	&	0.061	&	 REM                \\
60668.47	&	 r      	&	19.745	&	0.113	&	 LCO                \\
60670.75	&	 r      	&	20.434	&	0.071	&	 AFOSC              \\
60671.04	&	 r      	&	20.269	&	0.079	&	 REM                \\
60672.13	&	 r      	&	20.604	&	0.136	&	 REM                \\
60672.46	&	 r      	&	20.716	&	0.187	&	 LCO                \\
60673.15	&	 r      	&	20.823	&	0.244	&	 REM                \\
60674.13	&	 r      	&	21.016	&	0.253	&	 REM                \\
60675.06	&	 r      	&	20.867	&	0.204	&	 LCO                \\
60676.06	&	 r      	&	21.331	&	0.227	&	 REM                \\
60677.04	&	 r      	&	20.87	&	0.271	&	 LCO                \\
60680.09	&	 r      	&	21.785	&	0.311	&	 REM                \\
60684.12	&	 r      	&	20.129	&	0.14	&	 REM                \\

57710.38	&	 i      	&	 >21.008   	&	 -    	&	 PS1                \\
57971.58	&	 i      	&	 >22.599   	&	 -    	&	 PS1                \\
58063.39	&	 i      	&	 >22.614   	&	 -    	&	 PS1                \\
58719.55	&	 i      	&	 >22.659   	&	 -    	&	 PS1                \\
59101.6	&	 i      	&	 >22.458   	&	 -    	&	 PS1                \\
59511.41	&	 i      	&	 >21.654   	&	 -    	&	 PS1                \\
59838.48	&	 i      	&	 >22.060   	&	 -    	&	 PS1                \\
60220.55	&	 i      	&	 >21.469   	&	 -    	&	 PS1                \\
60630.34	&	 i      	&	 >19.873   	&	 -    	&	 PS1                \\
60642.98	&	 i      	&	16.936	&	0.017	&	 LCO                \\
60643.09	&	 i      	&	16.912	&	0.035	&	 REM                \\
60643.74	&	 i      	&	16.891	&	0.027	&	 67/91,Moravian-DIF \\
60644.11	&	 i      	&	16.869	&	0.083	&	 REM                \\
60644.81	&	 i      	&	16.873	&	0.014	&	 LCO                \\
60645.22	&	 i      	&	16.873	&	0.036	&	 REM                \\
60646.23	&	 i      	&	16.905	&	0.113	&	 REM                \\
60646.79	&	 i      	&	16.88	&	0.015	&	 LCO                \\
60648.03	&	 i      	&	16.977	&	0.029	&	 REM                \\
60649.08	&	 i      	&	17.049	&	0.014	&	 LCO                \\
60650.21	&	 i      	&	17.078	&	0.089	&	 REM                \\
60650.85	&	 i      	&	17.12	&	0.015	&	 LCO                \\
60652.23	&	 i      	&	17.188	&	0.103	&	 REM                \\
60654.13	&	 i      	&	17.287	&	0.064	&	 REM                \\
60656.07	&	 i      	&	17.447	&	0.115	&	 REM                \\
60656.46	&	 i      	&	17.514	&	0.028	&	 LCO                \\
60658.14	&	 i      	&	17.638	&	0.09	&	 REM                \\
60658.46	&	 i      	&	17.68	&	0.034	&	 LCO                \\
60660.46	&	 i      	&	17.855	&	0.033	&	 LCO                \\
60661.04	&	 i      	&	17.951	&	0.04	&	 REM                \\
60662.92	&	 i      	&	18.267	&	0.053	&	 LCO                \\
60663.12	&	 i      	&	18.362	&	0.133	&	 REM                \\
60664.46	&	 i      	&	18.461	&	0.038	&	 LCO                \\
60665.14	&	 i      	&	18.603	&	0.141	&	 REM                \\
60666.81	&	 i      	&	18.966	&	0.075	&	 LCO                \\
60667.16	&	 i      	&	18.985	&	0.088	&	 REM                \\
60668.47	&	 i      	&	19.583	&	0.099	&	 LCO                \\
60670.74	&	 i      	&	20.206	&	0.053	&	 AFOSC              \\
60671.04	&	 i      	&	19.92	&	0.133	&	 REM                \\ 
60672.13	&	 i      	&	20.086	&	0.18	&	 REM                \\
60672.47	&	 i      	&	20.391	&	0.161	&	 LCO                \\
60673.15	&	 i      	&	20.227	&	0.205	&	 REM                \\
60674.13	&	 i      	&	 >20.282   	&	 -    	&	 REM                \\
60675.06	&	 i      	&	20.345	&	0.155	&	 LCO                \\
60676.06	&	 i      	&	20.55	&	0.181	&	 REM                \\
60677.04	&	 i      	&	20.536	&	0.241	&	 LCO                \\
60680.09	&	 i      	&	20.918	&	0.281	&	 REM                \\
60684.12	&	 i      	&	 >19.880   	&	 -    	&	 REM                \\

60642.82	&	 z      	&	17.083  &	0.038 	&	 TTT                \\
60642.98	&	 z      	&	17.103	&	0.023	&	 LCO                \\
60642.99	&	 z      	&	16.993  &	0.033 	&	 TTT                \\
60643.09	&	 z      	&	17.013	&	0.145	&	 REM                \\
60643.96	&	 z      	&	17.055  &	0.033 	&	 TTT                \\
60644.11	&	 z      	&	16.98	&	0.276	&	 REM                \\
60644.81	&	 z      	&	16.964	&	0.024	&	 LCO                \\
60644.82	&	 z      	&	17.002  &	0.030 	&	 TTT                \\
60645.01	&	 z      	&	16.932  &	0.040 	&	 TTT                \\
60645.22	&	 z      	&	16.988	&	0.28	&	 REM                \\
60646.23	&	 z      	&	17.024	&	0.094	&	 REM                \\
60646.79	&	 z      	&	16.992	&	0.021	&	 LCO                \\
60646.82	&	 z      	&	17.005  &	0.195 	&	 TTT                \\
60646.99	&	 z      	&	17.058  &	0.038 	&	 TTT                \\
60647.82	&	 z      	&	17.040  &	0.036 	&	 TTT                \\
60647.99	&	 z      	&	16.978  &	0.035 	&	 TTT                \\
60648.03	&	 z      	&	17.086	&	0.17	&	 REM                \\
60649.08	&	 z      	&	17.091	&	0.021	&	 LCO                \\
60650.21	&	 z      	&	17.153	&	0.369	&	 REM                \\
60650.85	&	 z      	&	17.178	&	0.018	&	 LCO                \\
60652.23	&	 z      	&	17.23	&	0.31	&	 REM                \\
60654.13	&	 z      	&	17.356	&	0.171	&	 REM                \\
60656.07	&	 z      	&	17.505	&	0.16	&	 REM                \\
60656.46	&	 z      	&	17.508	&	0.033	&	 LCO                \\
60658.14	&	 z      	&	17.713	&	0.282	&	 REM                \\
60658.46	&	 z      	&	17.615	&	0.054	&	 LCO                \\
60660.46	&	 z      	&	17.87	&	0.052	&	 LCO                \\
60661.04	&	 z      	&	18.022	&	0.259	&	 REM                \\
60662.92	&	 z      	&	18.205	&	0.056	&	 LCO                \\
60663.12	&	 z      	&	18.261	&	0.158	&	 REM                \\
60664.46	&	 z      	&	18.357	&	0.049	&	 LCO                \\
60665.14	&	 z      	&	18.51	&	0.298	&	 REM                \\
60666.81	&	 z      	&	18.808	&	0.133	&	 LCO                \\
60667.16	&	 z      	&	18.806	&	0.48	&	 REM                \\
60668.48	&	 z      	&	19.224	&	0.117	&	 LCO                \\
60670.74	&	 z      	&	19.788	&	0.089	&	 AFOSC              \\
60671.04	&	 z      	&	 >19.383   	&	 -    	&	 REM                \\
60672.13	&	 z      	&	 >19.273   	&	 -    	&	 REM                \\
60672.47	&	 z      	&	19.957	&	0.209	&	 LCO                \\
60673.15	&	 z      	&	 >19.127   	&	 -    	&	 REM                \\
60674.13	&	 z      	&	 >19.122   	&	 -    	&	 REM                \\
60675.06	&	 z      	&	19.842	&	0.18	&	 LCO                \\
60676.06	&	 z      	&	 >19.333   	&	 -    	&	 REM                \\
60677.04	&	 z      	&	19.987	&	0.262	&	 LCO                \\
60680.09	&	 z      	&	 >19.488   	&	 -    	&	 REM                \\
60684.12	&	 z      	&	 >18.748   	&	 -    	&	 REM                \\

58345.6	&	 w      	&	 >22.788   	&	 -    	&	 PS1                \\
58351.59	&	 w      	&	 >22.488   	&	 -    	&	 PS1                \\
58489.24	&	 w      	&	 >22.189   	&	 -    	&	 PS1                \\
58733.54	&	 w      	&	 >22.372   	&	 -    	&	 PS1                \\
58751.49	&	 w      	&	 >23.444   	&	 -    	&	 PS1                \\
58752.5	&	 w      	&	 >23.212   	&	 -    	&	 PS1                \\
58813.3	&	 w      	&	 >23.675   	&	 -    	&	 PS1                \\
59105.52	&	 w      	&	 >23.576   	&	 -    	&	 PS1                \\
59140.45	&	 w      	&	 >23.527   	&	 -    	&	 PS1                \\
59195.28	&	 w      	&	 >22.591   	&	 -    	&	 PS1                \\
59218.24	&	 w      	&	 >23.225   	&	 -    	&	 PS1                \\
59222.29	&	 w      	&	 >22.081   	&	 -    	&	 PS1                \\
59465.56	&	 w      	&	 >23.300   	&	 -    	&	 PS1                \\
59486.55	&	 w      	&	 >22.440   	&	 -    	&	 PS1                \\
59490.4	&	 w      	&	 >23.236   	&	 -    	&	 PS1                \\
59500.38	&	 w      	&	 >22.242   	&	 -    	&	 PS1                \\
59516.32	&	 w      	&	 >23.048   	&	 -    	&	 PS1                \\
59527.28	&	 w      	&	 >22.359   	&	 -    	&	 PS1                \\
59543.29	&	 w      	&	 >23.391   	&	 -    	&	 PS1                \\
59810.61	&	 w      	&	 >23.168   	&	 -    	&	 PS1                \\
59822.54	&	 w      	&	 >23.647   	&	 -    	&	 PS1                \\
59827.53	&	 w      	&	 >23.037   	&	 -    	&	 PS1                \\
59850.49	&	 w      	&	 >23.315   	&	 -    	&	 PS1                \\
59870.38	&	 w      	&	 >23.046   	&	 -    	&	 PS1                \\
59907.37	&	 w      	&	 >22.642   	&	 -    	&	 PS1                \\
60175.59	&	 w      	&	 >22.886   	&	 -    	&	 PS2                \\
60208.51	&	 w      	&	 >22.024   	&	 -    	&	 PS1                \\
60223.46	&	 w      	&	 >22.438   	&	 -    	&	 PS1                \\
60227.44	&	 w      	&	 >22.674   	&	 -    	&	 PS2                \\
60232.39	&	 w      	&	 >22.321   	&	 -    	&	 PS2                \\
60235.39	&	 w      	&	 >22.977   	&	 -    	&	 PS1                \\
60238.4	&	 w      	&	 >22.771   	&	 -    	&	 PS2                \\
60256.34	&	 w      	&	 >22.888   	&	 -    	&	 PS1                \\
60259.38	&	 w      	&	 >22.408   	&	 -    	&	 PS1                \\
60264.33	&	 w      	&	 >22.434   	&	 -    	&	 PS1                \\
60283.31	&	 w      	&	 >23.031   	&	 -    	&	 PS2                \\
60290.23	&	 w      	&	 >22.820   	&	 -    	&	 PS1                \\
60559.55	&	 w      	&	 >22.170   	&	 -    	&	 PS2                \\
60563.51	&	 w      	&	 >23.368   	&	 -    	&	 PS1                \\
60564.55	&	 w      	&	 >23.306   	&	 -    	&	 PS1                \\
60581.5	&	 w      	&	 >22.857   	&	 -    	&	 PS2                \\
60588.42	&	 w      	&	 >22.261   	&	 -    	&	 PS1                \\
60592.45	&	 w      	&	 >22.380   	&	 -    	&	 PS2                \\
60605.39	&	 w      	&	 >22.555   	&	 -    	&	 PS2                \\
60635.33	&	 w      	&	19.876	&	0.028	&	 PS1                \\
60641.32	&	 w      	&	17.096	&	0.004	&	 PS1                \\
60642.98	&	 w      	&	16.877	&	0.1	&	 NOT                \\
60647.29	&	 w      	&	16.768	&	0.002	&	 PS2                \\

60616.07 & L      & >18.000   & -    & GOTO-L             \\
60636.92 & L      & 18.340   & 0.110 & GOTO-L             \\
60636.92 & L      & 18.335   & 0.108 & GOTO-L             \\
60640.48 & L      & 17.160   & 0.019 & GOTO-L             \\
60640.48 & L      & 17.277   & 0.043 & GOTO-L             \\
60656.54 & L      & 17.605   & 0.087 & GOTO-L             \\
60656.54 & L      & 17.684   & 0.069 & GOTO-L             \\
60664.86 & L      & 18.721   & 0.191 & GOTO-L             \\

60633.04 & q      & >21.030   & -    & BlackGem           \\
60640.04 & q      & 17.270   & 0.020 & BlackGem           \\

60608.99 & cyan   & >20.856   & -    & ATLAS              \\
60614.40 & cyan   & >20.695   & -    & ATLAS              \\
60616.99 & cyan   & >18.787   & -    & ATLAS              \\
60635.12 & cyan   & >20.717   & -    & ATLAS              \\
60636.34 & cyan   & 18.669   & 0.050 & ATLAS              \\
60639.11 & cyan   & 17.430   & 0.016 & ATLAS              \\
60640.35 & cyan   & 17.178   & 0.015 & ATLAS              \\
60642.98 & cyan   & 16.897   & 0.100 & NOT (spectrum)     \\
60643.12 & cyan   & 16.778   & 0.011 & ATLAS              \\
60644.32 & cyan   & 16.813   & 0.054 & ATLAS              \\
60647.09 & cyan   & 16.836   & 0.012 & ATLAS              \\
60672.80 & cyan   & 20.465   & 0.251 & ATLAS              \\
60696.26 & cyan   & >20.740   & -    & ATLAS              \\
60704.28 & cyan   & >19.673   & -    & ATLAS              \\

60602.47 & orange & >20.366   & -    & ATLAS              \\
60603.24 & orange & >19.912   & -    & ATLAS              \\
60604.41 & orange & >20.466   & -    & ATLAS              \\
60607.17 & orange & >20.613   & -    & ATLAS              \\
60611.19 & orange & >20.307   & -    & ATLAS              \\
60612.51 & orange & >19.998   & -    & ATLAS              \\
60615.18 & orange & >20.695   & -    & ATLAS              \\
60616.40 & orange & >20.655   & -    & ATLAS              \\
60618.27 & orange & >20.835   & -    & ATLAS              \\
60619.14 & orange & >20.634   & -    & ATLAS              \\
60620.90 & orange & >20.556   & -    & ATLAS              \\
60623.14 & orange & >20.520   & -    & ATLAS              \\
60624.53 & orange & >20.307   & -    & ATLAS              \\
60630.46 & orange & >19.749   & -    & ATLAS              \\
60631.15 & orange & >20.020   & -    & ATLAS              \\
60632.89 & orange & >20.384   & -    & ATLAS              \\
60634.35 & orange & >20.537   & -    & ATLAS              \\
60636.85 & orange & 18.265   & 0.044 & ATLAS              \\
60638.34 & orange & 17.740   & 0.016 & ATLAS              \\
60642.98 & orange & 16.923   & 0.100 & NOT (spectrum)     \\
60644.83 & orange & 16.847   & 0.015 & ATLAS              \\
60648.81 & orange & 16.983   & 0.022 & ATLAS              \\
60650.29 & orange & 17.083   & 0.015 & ATLAS              \\
60651.09 & orange & 16.829   & 0.018 & ATLAS              \\
60652.57 & orange & 17.205   & 0.017 & ATLAS              \\
60656.42 & orange & 17.680   & 0.161 & ATLAS              \\
60658.32 & orange & 17.823   & 0.045 & ATLAS              \\
60659.13 & orange & 17.791   & 0.083 & ATLAS              \\
60660.58 & orange & 17.964   & 0.036 & ATLAS              \\
60663.08 & orange & 18.381   & 0.051 & ATLAS              \\
60664.58 & orange & 18.621   & 0.063 & ATLAS              \\
60666.30 & orange & 18.982   & 0.071 & ATLAS              \\
60667.05 & orange & 19.099   & 0.120 & ATLAS              \\
60668.29 & orange & 19.686   & 0.154 & ATLAS              \\
60671.05 & orange & >20.257   & -    & ATLAS              \\
60672.30 & orange & 20.391   & 0.269 & ATLAS              \\
60674.27 & orange & 20.544   & 0.354 & ATLAS              \\
60678.26 & orange & 20.496   & 0.307 & ATLAS              \\
60679.04 & orange & >19.861   & -    & ATLAS              \\
60680.26 & orange & 20.675   & 0.358 & ATLAS              \\
60682.24 & orange & >20.126   & -    & ATLAS              \\
60684.26 & orange & >20.433   & -    & ATLAS              \\
60687.11 & orange & >19.309   & -    & ATLAS              \\
60688.80 & orange & >19.684   & -    & ATLAS              \\
60690.29 & orange & >19.915   & -    & ATLAS              \\
60692.82 & orange & >20.235   & -    & ATLAS              \\
60694.27 & orange & >20.433   & -    & ATLAS              \\

60643.09 & J      & 16.668   & 0.155 & REMIR              \\
60644.22 & J      & 16.609   & 0.367 & REMIR              \\
60646.23 & J      & 16.579   & 0.168 & REMIR              \\
60648.05 & J      & 16.603   & 0.095 & REMIR              \\
60650.24 & J      & 16.681   & 0.139 & REMIR              \\
60654.14 & J      & 16.809   & 0.100 & REMIR              \\
60656.15 & J      & 16.882   & 0.158 & REMIR              \\
60658.15 & J      & 16.945   & 0.139 & REMIR              \\
60661.06 & J      & 17.339   & 0.365 & REMIR              \\

60643.10 & H      & 16.327   & 0.212 & REMIR              \\
60644.23 & H      & 16.180   & 0.192 & REMIR              \\
60646.24 & H      & 16.120   & 0.118 & REMIR              \\
60648.05 & H      & 16.128   & 0.105 & REMIR              \\
60650.25 & H      & 16.163   & 0.258 & REMIR              \\
60654.15 & H      & 16.250   & 0.142 & REMIR              \\
60656.15 & H      & 16.312   & 0.146 & REMIR              \\
60658.16 & H      & 16.353   & 0.179 & REMIR              \\
60661.06 & H      & 16.548   & 0.177 & REMIR              \\

60643.11 & K      & 16.057   & 0.428 & REMIR              \\
60644.12 & K      & 15.917   & 0.273 & REMIR              \\
60645.23 & K      & 15.849   & 0.371 & REMIR              \\
60646.26 & K      & 15.778   & 0.307 & REMIR              \\
60648.03 & K      & 15.813   & 0.466 & REMIR              \\
60650.23 & K      & 15.851   & 0.374 & REMIR              \\
60652.24 & K      & 15.891   & 0.477 & REMIR              \\
60656.08 & K      & 15.951   & 0.458 & REMIR              \\
60658.22 & K      & 15.979   & 0.441 & REMIR              \\
60661.05 & K      & 16.108   & 0.348 & REMIR              \\

60641.84 & UVM2   & 16.650   & 0.080 & UVOT               \\
60643.67 & UVM2   & 16.530   & 0.050 & UVOT               \\
60645.70 & UVM2   & 16.640   & 0.050 & UVOT               \\
60650.99 & UVM2   & 18.060   & 0.090 & UVOT               \\
60656.42 & UVM2   & >18.970   & -    & UVOT               \\
60659.23 & UVM2   & >19.010   & -    & UVOT               \\

60641.84 & UVW2   & 17.010   & 0.080 & UVOT               \\
60643.67 & UVW2   & 16.860   & 0.060 & UVOT               \\
60645.70 & UVW2   & 17.060   & 0.060 & UVOT               \\
60650.99 & UVW2   & 18.250   & 0.110 & UVOT               \\
60656.42 & UVW2   & >19.540   & -    & UVOT               \\
60659.23 & UVW2   & >19.570   & -    & UVOT               \\

60641.84 & UVW1   & 16.430   & 0.100 & UVOT               \\
60643.67 & UVW1   & 16.120   & 0.090 & UVOT               \\
60645.70 & UVW1   & 16.350   & 0.090 & UVOT               \\
60650.99 & UVW1   & 17.360   & 0.120 & UVOT               \\
60656.42 & UVW1   & 18.670   & 0.210 & UVOT               \\
60659.23 & UVW1   & >18.730   & -    & UVOT               \\ \hline
\end{longtable}

\bsp	
\label{lastpage}
\end{document}